\documentclass[a4paper, amsfonts, amssymb, amsmath, onecolumn, showkeys, nofootinbib, superscriptaddress]{revtex4-1}

\usepackage{caption}
\usepackage{subcaption}
\usepackage{gensymb}
\usepackage{dcolumn}
\usepackage{bm}
\usepackage[pdftex]{graphicx,color} 
\usepackage{epstopdf}  

\begin{document}
\title{Enhanced absorption in thin and ultrathin silicon films by 3D photonic band gap back reflectors } 

\author{Devashish Sharma}
\affiliation{Complex Photonic Systems (COPS), 
MESA+ Institute for Nanotechnology, University of Twente, P.O. Box 217, 7500 AE Enschede, The Netherlands}
\affiliation{Mathematics of Computational Science (MACS), MESA+ Institute for Nanotechnology, University of Twente, P.O. Box 217, 7500 AE, Enschede, The Netherlands}
\affiliation{Present address: ASML Netherlands B.V., 5504 DR Veldhoven, The Netherlands}

\author{Shakeeb Bin Hasan}
\affiliation{Complex Photonic Systems (COPS), 
MESA+ Institute for Nanotechnology, University of Twente, P.O. Box 217, 7500 AE Enschede, The Netherlands}
\affiliation{Present address: ASML Netherlands B.V., 5504 DR Veldhoven, The Netherlands}

\author{Rebecca Saive}
\affiliation{Inorganic Materials Science (IMS), MESA+ Institute for Nanotechnology, University of Twente, P.O. Box 217, 7500 AE, Enschede, The Netherlands}

\author{Jaap J. W. van der Vegt}
\affiliation{Mathematics of Computational Science (MACS), MESA+ Institute for Nanotechnology, University of Twente, P.O. Box 217, 7500 AE, Enschede, The Netherlands}

\author{Willem L. Vos}
    \email{w.l.vos@utwente.nl, URL: www.photonicbandgaps.com}
\affiliation{Complex Photonic Systems (COPS), 
MESA+ Institute for Nanotechnology, University of Twente, P.O. Box 217, 7500 AE Enschede, The Netherlands}



\begin{abstract}
Since thin and ultrathin silicon films have limited optical absorption, we explore the effect of a nanostructured back reflector to recycle the unabsorbed light. 
As a back reflector, we investigate a three-dimensional (3D) photonic band gap crystal made from silicon that is readily integrated with the thin silicon films. 
We numerically obtain the optical properties by solving the 3D time-harmonic Maxwell equations using the finite-element method, and model silicon with experimentally determined optical constants. 
The absorption enhancement spectra and the photonic band gap generated current density are obtained by weighting the absorption spectra with the AM 1.5 standard solar spectrum. 
We study thin films in two different regimes, much thicker ($L_{Si} = 2400$ nm) or much thinner ($L_{Si} = 80$ nm) than the wavelength of light. 
At $L_{Si} = 2400$ nm thin film, the 3D photonic band gap crystal enhances the spectrally averaged ($\lambda = 680$ nm to $880$ nm) silicon absorption by $2.22\times$ ($s-$pol.) to $2.45\times$ ($p-$pol.), which exceeds the enhancement of a perfect metal back reflector ($1.47$ to $1.56 \times$). 
The absorption is considerably enhanced by the (i) broadband angle and polarization-independent reflectivity in the 3D photonic band gap, and (ii) the excitation of many guided modes in the film by the crystal's surface diffraction leading to greatly enhanced path lengths. 
At $L_{Si} = 80$ nm ultrathin film, the photonic crystal back reflector yields a striking average absorption enhancement of $9.15 \times$, much more than $0.83 \times$ for a perfect metal. 
This enhancement is due to a remarkable guided mode that is confined within the \textit{combined} thickness of the ultrathin film and the photonic crystal's Bragg attenuation length. 
An important feature of the 3D photonic band gap is to have a broad bandwidth, which leads to the back reflector's Bragg attenuation length being much shorter than the silicon absorption length. 
Consequently, light is confined inside the thin film and the remarkable absorption enhancements are not due to the additional thickness of the photonic crystal back reflector. We briefly discuss a number of high-tech devices that could profit from our results, including thin film solar cells.  
\end{abstract}


\maketitle

\section{Introduction}
\label{sect:Introduction}
Being a highly abundant and non-toxic material available in the earth's crust, silicon is an ideal choice to fabricate many high-tech devices with vast societal impact that employ the absorption of incident light. 
These devices include compact on-chip sensors, diodes and avalanche photodiodes, and charge-coupled devices (CCD) for cameras~\cite{Orton2004Book, Miller2009ProcIEEE, Reed2010NaturePhoton, Coteus2011IBM, Smit2012LPR, Vlasov2012IEEECommMag}. 
Moreover, to address the ongoing worldwide climate crisis~\cite{IPCC2021}, and provide solar energy for the entire world's population~\cite{Rohatgi1993JEM, IEA2020Report}, there is a pressing need to harvest the sun's energy with sustainable solar cells. 
All devices above employ the photovoltaic effect to absorb light and convert the absorbed energy into electricity using semiconductor materials~\cite{Bube1983Book, Luque2011Book}. 
While thick silicon devices are widely used, thin silicon films are enjoying a rising popularity on account of their obvious sustainability~\cite{Workshop2021OSA}, since they require less material and hence less resources and costs~\cite{Chapter11Luque2011Book}. 
Moreover, they are mechanically flexible so they can be deployed on many different platforms, including freely shaped ones. 

Since crystalline silicon (c-Si) has an indirect band gap at 1.1 eV, the absorption of light is low in the near infrared range that notably contains $36\%$ of all solar photons~\cite{ASTMG}. 
Conversely, the absorption length is long, namely $l_{a} = 1$ mm just above the gap at $\lambda = 1100$ nm (1.12 eV) and still only $l_{a} = 10~\mu$m at $\lambda = 800$ nm (1.55 eV)~\cite{Herzinger2998JAP}. 
Since the thickness of thin silicon film devices and solar cells is much less than the absorption lengths, the absorption of incident (solar) light is low~\cite{Sheng1983APL, Tiedje1984IEEETranscElectronDevices, Richter2013IEEEJournalofPhotovoltaics, Green2012ProgPhotovolt}, which adversely affects the cost and the flexibility advantages~\cite{Munzer1999IEEETransElectronDevices, Shah2004ProgPhotovolt, Feng2007IEEETransElectronDevices, YuPNAS2010}. 

\begin{figure}[ht]
\centering
\includegraphics[width=0.9\columnwidth]{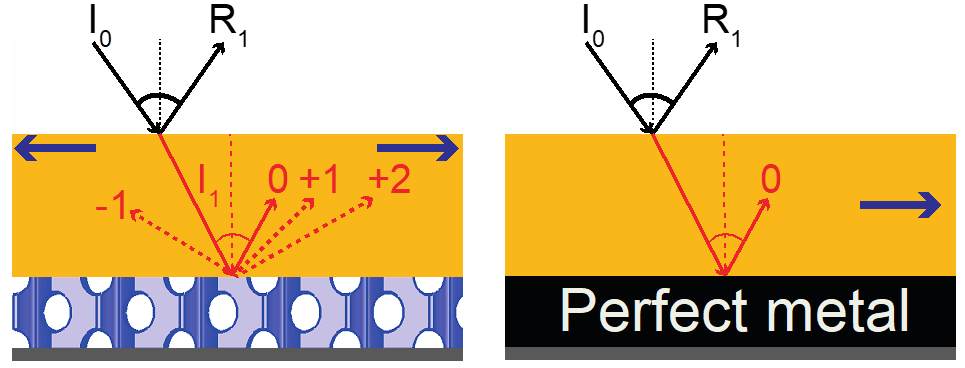}
\caption{Design of a thin silicon film (orange) with a 3D photonic band gap crystal (purple, left) and a perfect metal (black, right) as back reflectors. 
Here, $0$ indicates the $0^{\textrm{th}}$ diffraction order that corresponds to specular reflected light by the photonic crystal or the perfect metal. 
$I_{0}$ and $R_{1}$ represent the light incident and the first reflection at the front surface of the thin film, respectively. 
$I_{1}$ represents the light refracted into the thin film medium and incident on the photonic crystal. 
$-1$, $1$, and $2$ are nonzero diffraction orders. 
Blue arrows represent the propagation of guided modes in the thin silicon film. 
For potential application in light-absorbing high-tech devices (including photovoltaics), thin metallic rear contacts (grey) are sketched, but are not considered in our nanophotonic simulations. 
}
\label{fig:SolarCellBackReflector}
\end{figure}

Efficient light trapping enhances the absorption efficiency of silicon films while sustaining their advantages~\cite{Sheng1983APL, Feng2007IEEETransElectronDevices, Andreani2019AdvPhys, Saive2021ProgressinPV}. 
In traditional light trapping approaches as in solar cells, one increases the light paths using random texturing~\cite{Yablonovitch1982IEEETransED, Moulin2011EnergyProcedia} to scatter incident light into long oblique light paths and uses a back reflector to reflect unabsorbed light back into the thin film. 
In practice, perfect scattering is impossible to achieve, which limits the attainable efficiency~\cite{Brendel1996IEEETED}. 
An ideal back reflector reflects light incident from any angle, also referred to as omnidirectional reflectivity~\cite{Fink1998Science}, and ideally for all wavelengths and all polarizations of light. 
As illustrated in Fig.~\ref{fig:SolarCellBackReflector}(right), a perfect metal with $100\%$ reflectivity at all wavelengths and all polarizations would thus seem to be an ideal back reflector. 
In practice, no metal has $100\%$ reflectivity at all wavelengths due to Ohmic losses~\cite{Griffiths1998Book}. 
Moreover, light that is not reflected by a real metal gets absorbed, which produces heat and further limits the absorption efficiency of a thin film device.

Notably, much work has been devoted to light trapping strategies based on wave optics~\cite{John1984PRL, Muskens2008NanoLetters, Wang2010NanoLetters, Polman2012NatMater, Brongersma2014NatMat}. 
These strategies outperform random scattering optics approaches, typically over broad wavelength ranges, where one takes advantage of enhancements caused by constructive interference. 
To manipulate the interference, specially designed nano-structures are pursued, including 1D ("Bragg stack"), 2D, and 3D photonic crystals{\color{blue}~\cite{Nishimura2003JAmChemSoc, Mihi2005JPhysChemB, Colodrero2009AdvMater, Hsu2016OptComm, Branham2015AdvMater, Bermel2007OptExpress, OBrien2008AdvMater, Curtin2008ApplPhysLett, Zeng2008APL, Biswas2010SolarEnergyMatSolarCells, Wehrspohn2012JOpt, Sprafke2013OptExpress, Ishizaki2018JJAP}}. 

In this paper, we focus on photonic crystal back reflectors with a \textit{complete 3D photonic band gap}~\cite{Joannopoulos2008Book, Devashish2017PRB}, a frequency range for which the propagation of light is rigorously forbidden for all incident angles and all polarizations simultaneously, as recently demonstrated in experiments and calculations~\cite{Huisman2011PRB, Leistikow2011PRL, Yeganegi2014PRB, Hasan2018PRL, Grishina2019ACS, Adhikary2020OptExpress, Tajiri2020PRB, Mavidis2020PRB}.  
To illustrate this concept, Fig.~\ref{fig:SolarCellBackReflector} shows a schematic design of a thin silicon film (orange) with a 3D photonic band gap crystal back reflector (purple). 
Incident light with intensity $I_{0}$ is Fresnel diffracted to intensity $I_{1}$ within the thin film. 
When the incident light $I_{1}$ has a frequency in the photonic band gap, it is reflected by the photonic crystal~\cite{Joannopoulos2008Book} back into the film. 
The specular reflected beam corresponds to the $0^{\textrm{th}}$ diffraction order. 
Figure~\ref{fig:SolarCellBackReflector} illustrates non-zeroth order diffraction modes, \textit{e.g.}, $-1$, $1$, and $2$, that are generated at the periodic interface between the thin film and the photonic crystal. 
These diffraction modes couple light into guided modes that are confined inside the thin silicon film via total internal reflection. 
Consequently, guided photons obtain a long path length inside the thin film and have thus an enhanced probability for absorption. 
Hence, different from a perfect metal, a 3D photonic crystal enhances the absorption of a thin silicon film by (i) profiting from perfect reflectivity inside the band gap for all incident angles and polarizations, and (ii) by generating guided modes~\cite{OBrien2008AdvMater, Biswas2010SolarEnergyMatSolarCells}. 

Recently, we reported a numerical study on the enhanced energy density and optical absorption for realistic and finite 3D silicon photonic band gap crystals with an embedded resonant cavity~\cite{Devashish2019PRB, Hack2019PRB}. 
The absorption was found to be substantially enhanced, but only within the tiny cavity volume, as opposed to the present case where the absorption occurs throughout the whole film volume, which avoids local heating and non-linear or many-body effects that adversely affect the absorption efficiency~\cite{Luque2011Book}. 
While a structure with a 3D photonic band gap is from the outset relevant as an omnidirectional, broadband, and polarization-robust back reflector for ultrathin silicon films (including solar cells), they have been hardly studied before. 
Therefore, we investigate nanostructured back reflectors with a 3D photonic band gap that is tailored to have a broad photonic band gap in the visible regime. 
Using numerical finite-element solutions of the 3D time-harmonic Maxwell equations, we calculate the absorption of light in a thin silicon film with a 3D inverse woodpile photonic crystal as a back reflector. 
To make our calculations relevant to experimental studies, we employ a dispersive and complex refractive index obtained from experiments~\cite{Green2008SolarEnergyMater} and compare the photonic crystal back reflector to a perfect metallic back reflector. 
We verify that the absorption is not enhanced by the extra material volume. 
Ultimately, we aim to understand the physics behind large enhancements by identifying the relevant physical mechanisms compared to a standard back reflector.

\section{Methods}
\label{sect:Methods}
\subsection{Structure}
\label{subsect:Structure}

For the photonic band gap back reflector we have chosen the cubic inverse woodpile crystal structure~\cite{Ho1994SSC}, on account of its broad band gap that is robust to disorder~\cite{Hillebrand2003JAP, Woldering2009JAP}, and since this structure is readily fabricated from silicon~\cite{vandenBroek2012AFM}, and thus a suitable candidate for integration with thin silicon film devices. 
The crystal structure is shown in Fig.~\ref{fig:ComputationalCell}(a) and consists of two arrays of identical nanopores with radius $r$ running in two orthogonal directions $\mathrm{X}$ and $\mathrm{Z}$. 
Each nanopore array has a centered-rectangular lattice with lattice parameters $c$ and $a$, see also Appendix~\ref{sect:VolumeFractionInverseWoodpile}. 
For a ratio $\frac{a}{c} = \sqrt{2}$, the diamond-like structure has cubic symmetry. 
Cubic inverse woodpile photonic crystals have a broad maximum band gap width $\Delta \omega / \omega_{c} = 25.3\%$ relative to the central band gap frequency $\omega_{c}$ for pores with a relative radius $\frac{r}{a} = 0.245$~\cite{Hillebrand2003JAP, Woldering2009JAP}. 
Our prior results reveal that a reflectivity in excess of $R > 99\%$ and a transmission $T < 1 \%$ occur already for a thin inverse woodpile photonic crystal with a thickness of a few unit cells ($L_{3DPC} \geq 3c$)~\cite{Devashish2017PRB, Tajiri2020PRB}. 
Therefore, we choose here a cubic inverse-woodpile crystal with an optimal pore radius $\frac{r}{a} = 0.245$ and with a thickness $L_{3DPC} = 4c = 1200$ nm as a back reflector for the calculation of the absorption of light by the thin silicon film. 

\subsection{Computations}
\label{subsect:Computation}

\begin{figure}[ht]
\centering
\begin{subfigure}[b]{0.9\textwidth}
     \includegraphics[width=0.7\columnwidth]{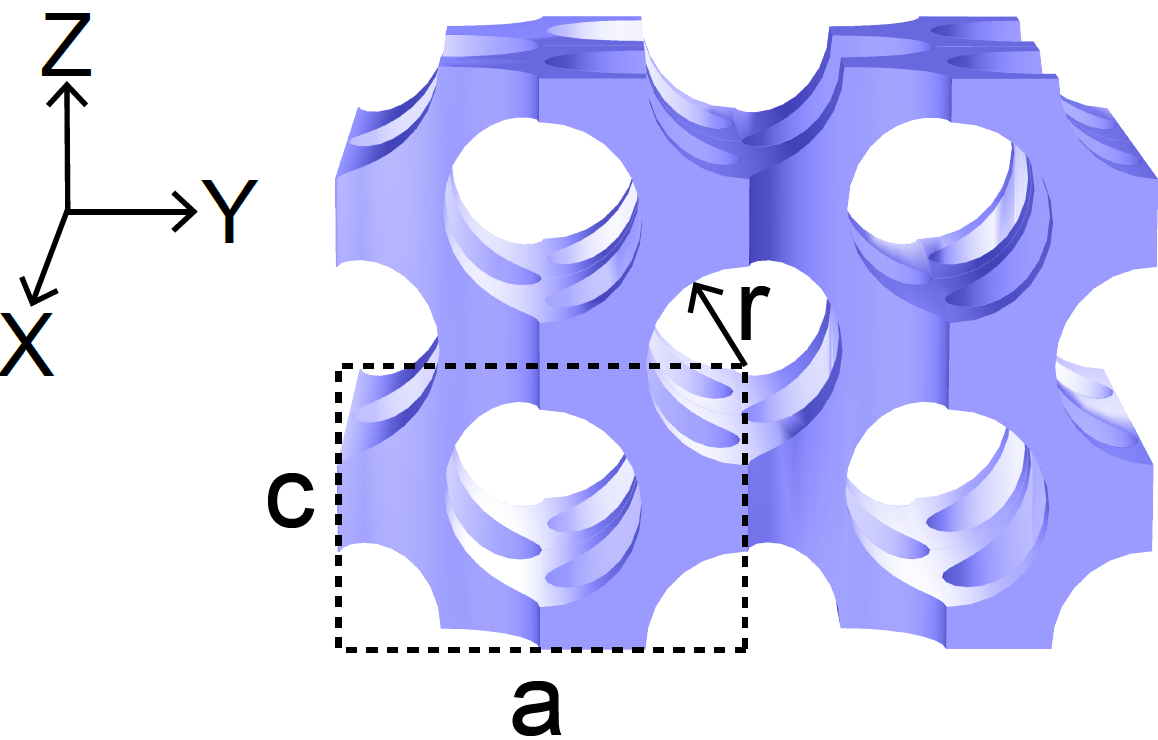}
     \caption{}
\end{subfigure}
\hfill
\begin{subfigure}[b]{0.9\textwidth}
    \includegraphics[width=0.9\columnwidth]{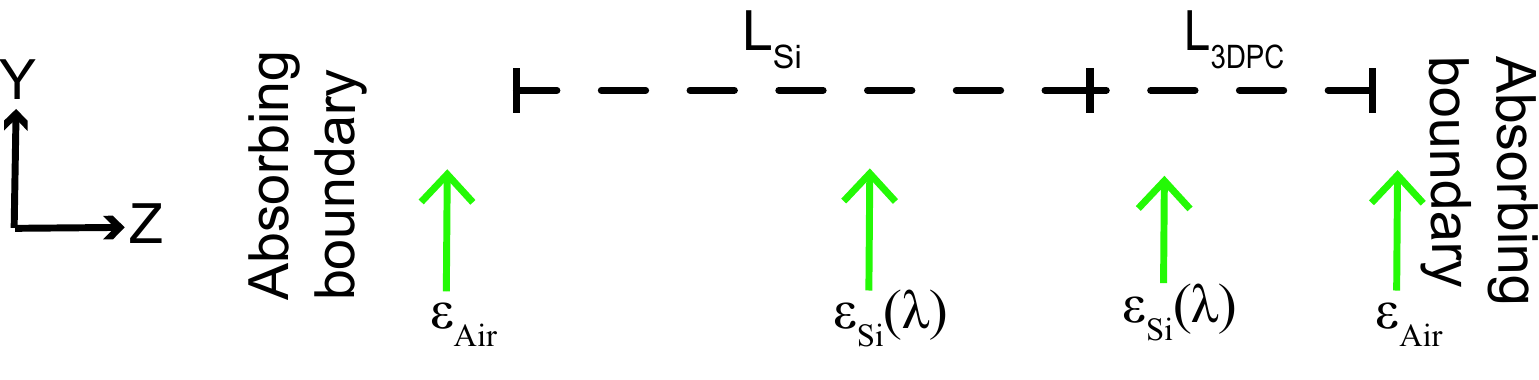}
    \caption{}
\end{subfigure}
\caption{
(a) Schematic of the 3D inverse woodpile photonic crystal structure with the $\mathrm{XYZ}$ coordinate axes. 
We show a $2\times2\times2$ supercell, with two arrays of identical nanopores with radius $r$ parallel to the $\mathrm{X}$ and $\mathrm{Z}$ axes. 
The lattice parameters of the tetragonal unit cell are $c$ and $a$, in a ratio $\frac{a}{c} = \sqrt{2}$ for cubic symmetry. 
The blue color represents the high-index material with a dielectric function similar to silicon. 
(b) Computational cell bounded by absorbing boundaries at $-\mathrm{Z}$ and $+\mathrm{Z}$, and by periodic boundary conditions at $\pm X$ and $\pm Y$. 
The thin silicon film with thickness $L_{Si} = 2400$ nm is the absorbing layer and a 3D inverse woodpile photonic crystal with thickness $L_{3DPC} = 4c = 1200$ nm is the back reflector.}
\label{fig:ComputationalCell}
\end{figure}

To calculate the optical absorption in a thin silicon film, we employ the commercial COMSOL Multiphyiscs finite-element (FEM) software to solve the time-harmonic Maxwell equations~\cite{COMSOLMultiphysics}. 
Figure~\ref{fig:ComputationalCell} (b) illustrates the computational cell viewed in the $YZ$ plane. 
The incident fields start from a plane at the left that is separated from the silicon layer by an air layer. 
Since the plane also absorbs the reflected waves~\cite{Jin1993Book}, it represents a boundary condition rather than a true current source. 
The incident plane waves have either $s$ polarization (electric field normal to the plane of incidence) or $p$ polarization (magnetic field normal to the plane of incidence), and have an angle of incidence between $\theta = 0^{\circ}$ and $80^{\circ}$. 
We employ Bloch-Floquet periodic boundaries in the $\pm X$ and the $\pm Y$ directions to describe the infinitely extended thin silicon film~\cite{Joannopoulos2008Book}. 
To describe a thin film with finite support, absorbing boundaries are employed in the $-Z$ and $+Z$ directions. 
We calculate reflectivity and transmission of the thin film at the absorbing boundaries in the $-Z$ and $+Z$ directions, respectively. 
The light with a given wavelength $\lambda$ incident at an angle $\theta$ with respect to the surface normal is either reflected or transmitted, or absorbed by the thin film~\cite{Griffiths1998Book}. 
To calculate the absorption $A_{Si}(\lambda, \theta)$ of a thin silicon film, we employ the relation 
\begin{equation}
\label{eq:absorption}
A_{Si}(\lambda, \theta) = 1 - R_{Si}(\lambda, \theta) - T_{Si}(\lambda, \theta),
\end{equation}
with $R_{Si}(\lambda, \theta)$ the reflectivity and $T_{Si}(\lambda, \theta)$ the transmission spectra that are normalized to the incident light intensity $I_{0}$. 

To gauge the performance of a 3D photonic band gap back reflector, we define the wavelength and angle-dependent absorption enhancement $\eta_{abs}(\lambda, \theta)$ of the thin silicon film in Fig.~\ref{fig:ComputationalCell} as
\begin{equation}
\eta_{abs}(\lambda, \theta) \equiv  \frac{\int^{\lambda+\Delta \lambda}_{\lambda-\Delta \lambda} {P_{AM 1.5}(\lambda)} A_{(Si+3DPC)}(\lambda,\theta) d\lambda}{\int^{\lambda+\Delta \lambda}_{\lambda-\Delta \lambda} {P_{AM 1.5}(\lambda)} A_{Si}(\lambda,\theta) d\lambda},
\label{eq:EnhancementRatio}
\end{equation}
where the absorption is weighted with the solar spectrum using the air mass coefficient $P_{AM 1.5}(\lambda)$~\cite{ASTMG}. 
In Eq.~\ref{eq:EnhancementRatio}, $A_{(Si+3PC)}$ represents the absorption in a thin silicon film with a 3D photonic crystal back reflector and $A_{Si}$ the absorption in a thin silicon film with the same thickness, yet no back reflector. 
Using Eq.~\ref{eq:EnhancementRatio}, the enhancement is averaged over a bandwidth $(2 \Delta \lambda)$  at every discrete wavelength $\lambda$. 

At normal incidence ($\theta = 0^\circ$) the absorption enhancement $\eta^{'}_{abs}(\lambda)$ is deduced from Eq.~\ref{eq:EnhancementRatio} to 
\begin{equation}
\eta^{'}_{abs}(\lambda) \equiv \frac{\int^{\lambda+\Delta \lambda}_{\lambda-\Delta \lambda} {P_{AM 1.5}(\lambda)} A_{(Si+3DPC)}(\lambda) d\lambda}{\int^{\lambda+\Delta \lambda}_{\lambda-\Delta \lambda} {P_{AM 1.5}(\lambda)} A_{Si}(\lambda) d\lambda}. 
\label{eq:NormalEnhancementRatio}
\end{equation}
To calculate the angle-averaged absorption enhancement $\eta^{''}_{abs}(\lambda)$, the enhancement $\eta_{abs}(\lambda, \theta)$ (Eq.~\ref{eq:EnhancementRatio}) is averaged over $n$ incident angles $\theta_{i}$ to 
\begin{equation}
\eta^{''}_{abs}(\lambda) \equiv \frac{1}{n}\sum_{i = 1}^{i = n}{\eta_{abs}(\lambda, \theta_{i})}.
\label{eq:AllAnglesEnhancementRatio}
\end{equation}

Figure~\ref{fig:ComputationalCell} (b) illustrates the finite element mesh of tetrahedra that are used to subdivide the 3D computational cell into elements~\cite{ComputationalMesh}. 
Since the computations are intensive due to a finite element mesh of 167000 tetrahedra, we performed the calculations on the powerful ``Serendipity" cluster~\cite{Serendipity} at MACS in the MESA+ Institute (see also Ref.~\cite{Devashish2019PRB}). 

To enhance the weak absorption of silicon above the electronic band gap at wavelengths in the range $600$ nm $< \lambda < 1100$ nm, we tailor the lattice parameters of the inverse woodpile photonic crystal to $a = 425$ nm and $c = 300$ nm such that the band gap is in the visible range. 
The chosen lattice parameters are $37 \%$ smaller than the ones usually taken for photonic band gap physics in the telecom range~\cite{Devashish2017PRB, Hasan2018PRL, Grishina2019ACS, Adhikary2020OptExpress, Tajiri2020PRB, Mavidis2020PRB, Huisman2011PRB, Leistikow2011PRL, Uppu2021PRL}. 
The nanopore radii are taken to be $\frac{r}{a} = 0.245$, i.e., $r = 104$ nm in order for the broadest possible band gap~\cite{Hillebrand2003JAP, Woldering2009JAP}. 
The required dimensions are well within the feasible range of nanofabrication parameters~\cite{Tjerkstra2011JVSTB, vandenBroek2012AFM, Grishina2015NT}. 

To benchmark our proposition of using a 3D inverse woodpile photonic crystal as a back reflector, we compare the absorption spectra to spectra for the same thin silicon layer with a perfect and omnidirectional metallic back reflector. 
Therefore, in our simulations we replace the photonic crystal and the air layer on the right in Fig.~\ref{fig:ComputationalCell} (b) with a homogeneous metallic plane with a large and purely imaginary refractive index $n^{''} = -i \cdot 10^{20}$. 

For reference, we also show results for ideal Lambertian scattering~\cite{Yablonovitch1982JOSA}, with the understanding that in our calculations we do not consider a scattering front (or back) surface. 
In the Lambertian case, the absorption $A^{L}_{Si}(\lambda)$ of a thin silicon film with thickness $L_{Si}$ and refractive index $n_{Si}(\lambda) = \mathbb{R}\mathrm{e}(n_{Si} (\lambda)) + \mathbb{I}\mathrm{m}(n_{Si} (\lambda))$ is equal to~\cite{Green2002ProgPhotovoltaics, Massiot2020NatEnergy} 
\begin{equation}
A^{L}_{Si}(\lambda) = \frac{\alpha(\lambda) L_{Si}}{\alpha(\lambda) L_{Si} + \frac{1}{F(\lambda)}} , 
\label{eq:LambertianAbsorption}
\end{equation}
where $\alpha(\lambda)$ is the absorption coefficient that is equal to $\alpha(\lambda) = \frac{4\pi\mathbb{I}\mathrm{m}(n_{Si}(\lambda))}{\lambda}$, and $F(\lambda)$ the optical path length enhancement factor equal to $F(\lambda) = 4~{\mathbb{R}\mathrm{e}(n_{Si}(\lambda))^2}$. 

\section{Results and discussion} 
In the first two subsections, we show the main results, namely the absorption enhancements for silicon films thicker and thinner than the wavelength. 
In the subsequent subsections, we discuss detailed physical backgrounds, including film thickness, angular acceptance, and absence of photonic crystal backbone contributions to the overall absorption. 

\subsection{Supra-wavelength silicon film} 
\label{subsec:ComparisonPerfectMetal}

\begin{figure}[h]
\centering
\includegraphics[width=0.75\columnwidth]{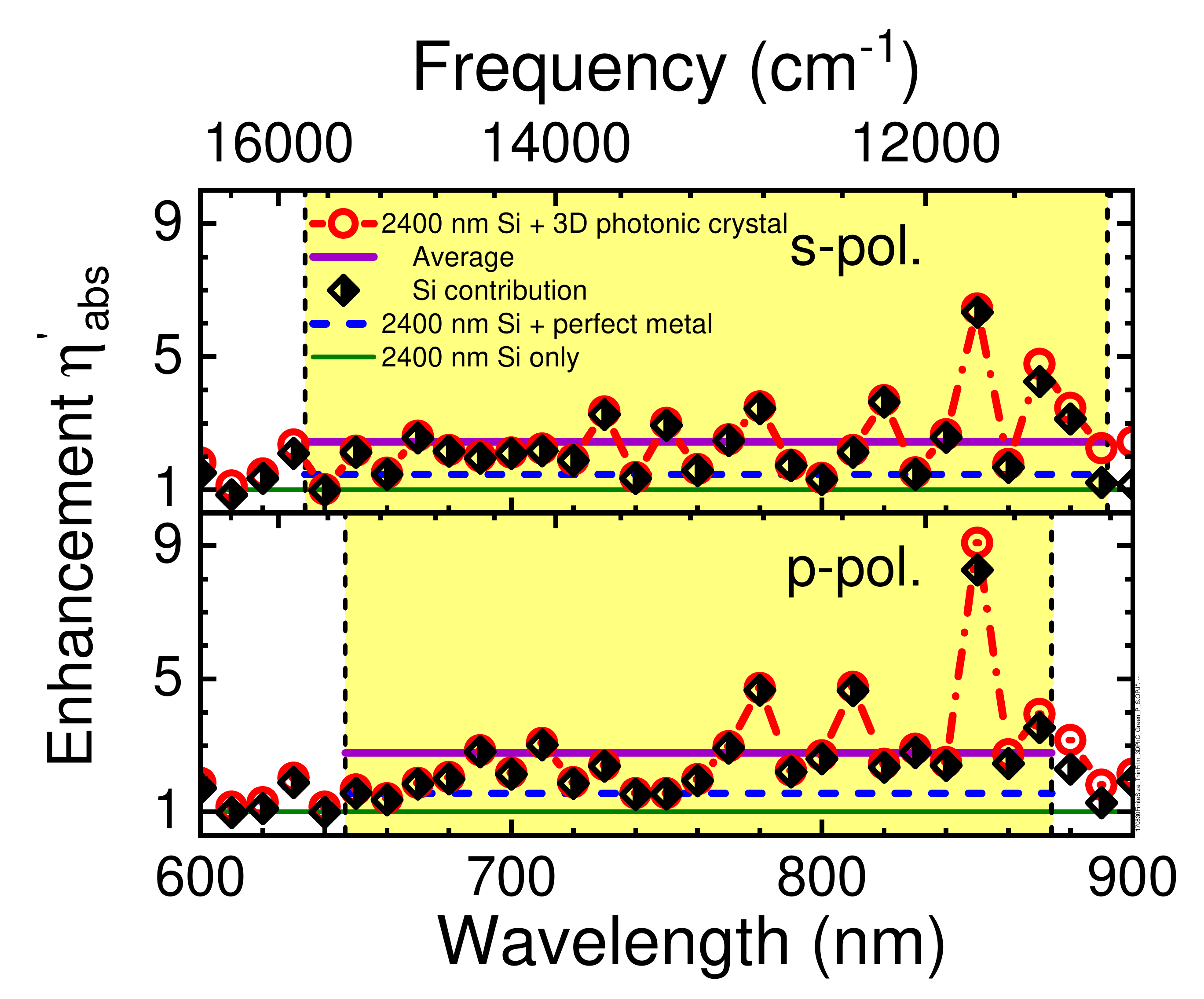}
\caption{Normal incidence absorption enhancement spectra $\eta^{'}_{abs}(\lambda)$ for a thin silicon film ($L_{Si} = 2400$ nm) with several back reflectors computed using Eq.~\ref{eq:NormalEnhancementRatio} with 
a bandwidth $2 \Delta \lambda = 10$ nm at each wavelength bin $\lambda$. 
The top panel is for $s-$polarized light and the bottom panel for $p-$polarized light. 
Red connected circles are data for a thin film with the 3D photonic crystal back reflector, the purple horizontal line is the average enhancement (using Eq.~\ref{eq:NormalEnhancementRatio} with a bandwidth $2 \Delta \lambda$ equal to the $s-$ and $p-$stop bands, respectively), the black diamonds are for the thin film part only, and the blue short-dashed line is for the thin film with a perfect metal back reflector. 
The vertical dashed lines show the edges of the $s-$ and $p-$stop bands, with the stop bands shown as yellow bars. 
}
\label{fig:EnhancementPS}
\end{figure}

Figure~\ref{fig:EnhancementPS} shows the normal incidence absorption enhancement $\eta^{'}_{abs}(\lambda)$ (Eq.~\ref{eq:NormalEnhancementRatio}) of a supra-wavelength $L_{Si} = 2400$ nm thin silicon film for both the perfect metal and the 3D photonic crystal back reflectors. 
For the 3D inverse woodpile back reflector, the absorption enhancement varies between  $\eta^{'}_{abs} = 1 \times$ and $9 \times$ inside the stop bands between $\lambda = 640$ nm and $900$ nm. 
The wavelength-averaged absorption enhancement is about $\langle\eta^{'}_{abs}\rangle$ $= 2.22 \times$ for the $s-$stop band and $\langle\eta^{'}_{abs}\rangle$ $= 2.45 \times$ for the $p-$stop band. 
In comparison, for a perfect metal back reflector the wavelength-averaged absorption enhancement is about $\langle\eta^{'}_{abs}\rangle$ $= 1.47\times$ for the $s-$stop band and $\langle\eta^{'}_{abs}\rangle$ $= 1.56\times$ for the $p-$stop band. 
Since a perfect metal back reflector has $100\%$ specular reflectivity only in the specular $0^{\textrm{th}}$ diffraction order, the absorption enhancement $\eta^{'}_{abs}$ is always less than two: $\langle\eta^{'}_{abs}\rangle  \leq 2$. 
In contrast, a photonic band gap crystal back reflector also has non-zero order diffraction modes, see Fig.~\ref{fig:SolarCellBackReflector}, that scatter light into guided modes where light is confined inside the thin silicon film via total internal reflection. 
Since the effective optical path length travelled by a photon in a guided mode is longer than the path length travelled with only the $0^{\textrm{th}}$ order diffraction mode, a photonic crystal back reflector yields a greater absorption enhancement $\langle\eta^{'}_{abs}\rangle$ $\geq 2$ for the diffracted wavelengths, as is apparent in Fig.~\ref{fig:EnhancementPS}.  
This observation is a first support of the notion that a 3D photonic band gap back reflector enhances the absorption of a thin silicon film by (i) behaving as a perfect reflector with nearly $100\%$ reflectivity for both polarizations, and (ii) exciting guided modes within the thin film. 

Since the high-index backbone of the 3D inverse woodpile photonic crystal consists of silicon, one might surmise that the absorption is enhanced by the addition of the photonic crystal's silicon backbone to the thin silicon film. 
To test this hypothesis, we calculate the absorption enhancement within the thickness $L_{Si}$ of the thin silicon film part (see Fig.~\ref{fig:ComputationalCell} (b)) using the volume integral of the total power dissipation density from Ref.~\cite{COMSOLMultiphysics}. 
Figure~\ref{fig:EnhancementPS} shows that the absorption enhancement spectra for the thin film volume agree very well with the spectra of the complete device (both thin film and photonic crystal) in the stop bands between $\lambda = 640$ nm and $900$ nm. 
Therefore, the high-index backbone of the 3D photonic crystal contributes \textit{negligibly} to the absorption inside the stop bands, even in the visible regime. 
Apparently, light that travels from the thin film into the photonic crystal is reflected back into the thin film, even before it is absorbed in the photonic crystal. 
To bolster this conclusion, we start with the notion that the typical length scale for reflection by a photonic crystal is the Bragg attenuation length $\ell_{Br}$~\cite{Vos2015Cambridge} that qualitatively equals the ratio of the central stop band wavelength and the photonic interaction strength $S$ times $\pi$: $\ell_{Br} = \lambda_c / (\pi~S)$~\cite{Vos2001NATO, Vos1996PRB}, where the strength $S$ is gauged by the ratio of the dominant stop band width and the central wavelength ($S = \Delta \lambda / \lambda_c$). 
From the stop band between $\lambda = 640$ nm and $900$ nm ($S = 260/770 = 0.34$) we arrive at $\ell_{Br} = 770/(\pi~0.34) = 725$ nm, which is much less than the absorption length of silicon: $\ell_{Br} << l_{a}$. 
Thus, the broad bandwidth of the back reflector's 3D photonic band gap is an important feature to enhance the absorption inside the thin silicon film itself, as the broad bandwidth corresponds to a short Bragg length.

\subsection{Sub-wavelength silicon film} 
\label{subsec:Subwavelength}

\begin{figure}[ht]
\centering
\includegraphics[width=0.75\columnwidth]{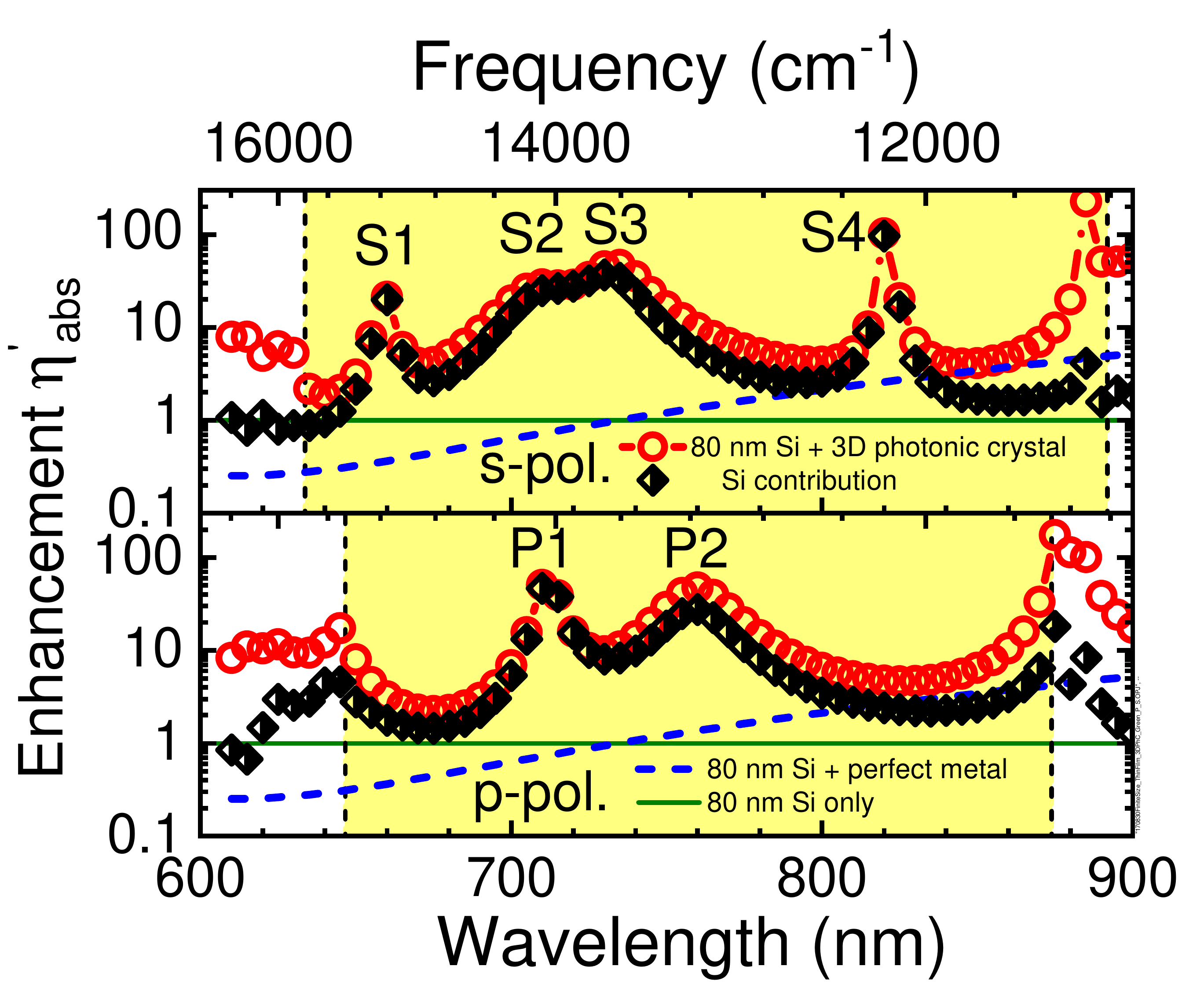} 
\caption{
Absorption enhancement spectra $\eta^{'}_{abs}(\lambda)$ for a sub-wavelength ultrathin silicon layer ($L_{Si} = 80$ nm), taken as ratio of absorption $A$ with a back reflector and absorption $A_{Si}$ without back reflector, all at normal incidence computed using Eq.~\ref{eq:NormalEnhancementRatio}. 
Top panel: $s$-polarized, bottom panel: $p$-polarized light, with $s-$ and $p-$stop bands shown as yellow highlighted regions. 
Red connected circles pertain to the ultrathin film with a 3D photonic crystal back reflector and the black diamonds are for the silicon part only. 
The blue dashed curves pertain to the ultrathin film with a perfect metal back reflector and the green lines are the reference level of the ultrathin film without a back reflector.
}
\label{fig:SubwavelengthPS}
\end{figure}

\begin{figure}[ht]
\centering
\includegraphics[width=0.75\columnwidth]{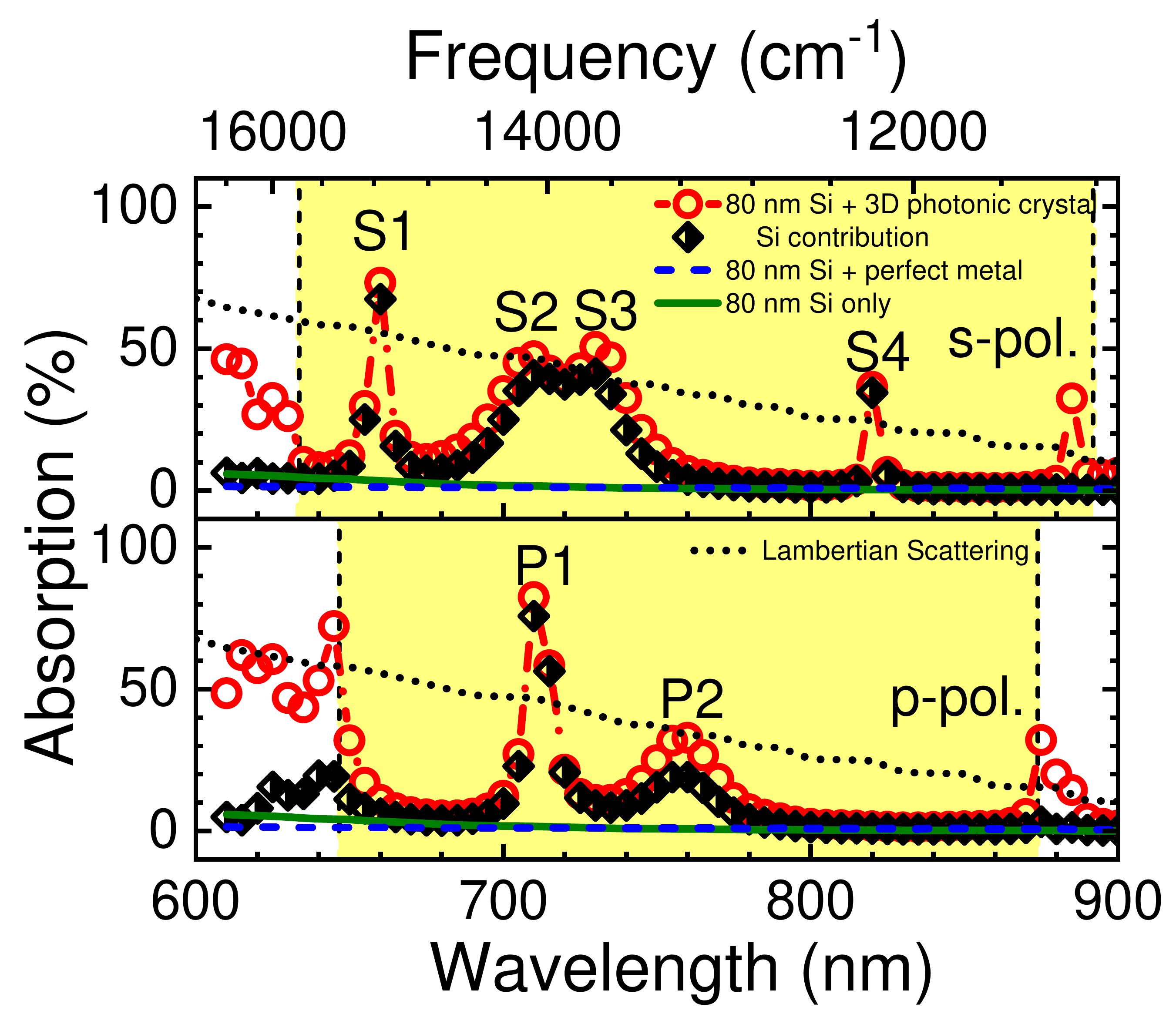} 
\caption{
Absolute absorption spectra (in $\%$) for a sub-wavelength ultrathin silicon layer ($L_{Si} = 80$ nm) at normal incidence. 
Top panel: $s-$polarized, bottom panel: $p-$polarized light, with $s-$ and $p-$stop bands shown as yellow highlighted regions. 
Red connected circles pertain to the ultrathin film with a 3D photonic crystal back reflector, whereas the black diamonds show the contribution of the ultrathin film only. 
Green solid curves pertain to an ultrathin film without a back reflector, and blue dashed curves to an ultrathin film with a perfect metal back reflector. 
Black dotted curves represent absorption $A^{L}_{Si}(\lambda)$ with Lambertian scattering, from Eq.~\ref{eq:LambertianAbsorption}.
}
\label{fig:SubwavelengthPS_Absorption}
\end{figure}

Figure~\ref{fig:SubwavelengthPS} shows the normal incidence absorption enhancement $\eta^{'}_{abs}(\lambda)$ (Eq.~\ref{eq:NormalEnhancementRatio}) of a $L_{Si} = 80$ nm ultrathin silicon film, whose thickness is much less than the wavelength in the material ($L_{Si} << \lambda / n$), hence no guided modes are sustained in the film itself. 
Since the thickness is also much less than the silicon absorption length $L_{Si} << l_{a} = 1000$ nm between $\lambda = 600$ nm and $900$ nm~\cite{Green2008SolarEnergyMater}, the absorption $A_{Si}$ of the ultrathin film itself is low, namely about $4.5 \%$ at the blue edge of the photonic stop bands) and $0.2 \%$ at the red edge of the stop band, see Fig.~\ref{fig:SubwavelengthPS_Absorption}. 
These results agree with a Fabry-P\'erot analysis of the ultrathin film, which reveals a broad first order resonance near $\lambda = 570$ nm that explains the increased absorption towards the short wavelengths. 

For the perfect metal back reflector in Fig.~\ref{fig:SubwavelengthPS}, the absorption is reduced ($\eta^{'}_{abs}(\lambda) < 1$) at wavelengths $\lambda < 730$ nm and enhanced ($\eta^{'}_{abs}(\lambda) > 1$) at longer wavelengths. 
This result is understood from the Fabry-P\'erot behavior of the ultrathin film in presence of the perfect metal, that induces an additional $\pi$ phase shift (due to the exit surface reflectivity) to the round-trip phase. 
Consequently, a first order resonance appears near $\lambda = 1140$ nm, and the next anti-resonance near $\lambda = 600$ nm. 
Therefore, the absorption is enhanced towards $\lambda = 1140$ nm and reduced towards $\lambda = 600$ nm, in agreement with the results in Fig.~\ref{fig:SubwavelengthPS}. 
The corresponding wavelength-averaged absorption enhancements are $\langle \eta^{'}_{abs} \rangle$ $= 0.8\times$ for $s-$polarized and $\langle\eta^{'}_{abs}\rangle$ $= 0.85\times$ for the $p-$polarized stop band. 
In other words, a perfect metal back reflector hardly enhances the absorption for ultrathin films with the thickness corresponding to a Fabry-P\'erot minimum for the desired wavelengths of absorption enhancement. 

In presence of a photonic band gap back reflector, the wavelength-averaged absorption enhancement is surprisingly larger, see Fig.~\ref{fig:SubwavelengthPS}. 
The strong increase is clearly illustrated by the use of a logarithmic scale. 
We see several peaks between $\lambda = 600$ nm and $900$ nm in Fig.~\ref{fig:SubwavelengthPS}, four absorption resonances (S1, S2, S3, and S4) inside the $s-$polarized stop band and two absorption resonances (P1 and P2) inside the $p-$polarized stop band with enhancements as high as $\eta^{'}_{abs} = 100\times$ for both polarizations. 
The wavelength-averaged enhancements are $\langle\eta^{'}_{abs}\rangle$ $= 13.5\times$ for the $s-$polarized stop band and $\langle\eta^{'}_{abs}\rangle$ $= 11.4\times$ for the $p-$polarized stop band. 
Since the thickness of the ultrathin silicon $L_{Si} = 80$ nm does not sustain guided modes at the wavelengths within these stop bands, the enhanced absorption peaks must be induced by the presence of the photonic crystal back reflector, which has an effective refractive index smaller than that of silicon (see Appendix~\ref{sect:VolumeFractionInverseWoodpile}), and hence leads to no phase shift like the metallic back reflector. 

Considering that this ultrathin silicon layer is much thinner than the one in Sec.~\ref{subsec:ComparisonPerfectMetal}, there is a likelihood that the absorption is enhanced by the extra silicon volume from the photonic crystal's backbone, whose thickness is much greater than the film thickness, namely $L_{3DPC} = 1200$ nm versus $L_{Si} = 80$ nm. 
To test this hypothesis, we calculated the absorption within the volume of the ultrathin silicon film part only ($L_{Si} = 80$ nm), using the power per unit volume formula of Ref.~\cite{COMSOLMultiphysics}, as shown in Fig.~\ref{fig:SubwavelengthPS_Absorption} (from which Fig.~\ref{fig:SubwavelengthPS} is deduced) with various back reflectors for $s$ and $p$ polarizations. 
The enhanced absorption at all six resonances S1-S4, and P1-P2 occur in the ultrathin film volume only. 
The wavelength-averaged absorption enhancement within the ultrathin film part is $\langle\eta^{'}_{abs}\rangle = 10.46\times$ averaged over the $s-$polarized stop band and $\langle\eta^{'}_{abs}\rangle = 7.84\times$ for the $p-$polarized stop band, which are nearly the same as in the full device. 
Therefore, we conclude that the absorption in the photonic crystal hardly contributes to the enhanced absorption and does not induce the presence of the absorption peaks. 
Figure~\ref{fig:SubwavelengthPS} reveals that the absorption within the ultrathin film and the one within the whole device (both ultrathin film and photonic crystal) match very well near the center of the $s-$ and $p-$stop bands and differ at the edges. 
The reason is that the Bragg attenuation length $\ell_{Br}$ is smallest near the center of the stop band, while it increases toward the band edges where it leads to more absorption in the crystal before the light is reflected within a Bragg length. 

For reference, Fig.~\ref{fig:SubwavelengthPS_Absorption} also shows the absorption limit due to Lambertian scattering (Eq.~\ref{eq:LambertianAbsorption}) for the $L_{Si} = 80$ nm ultrathin film. 
We note that in between the resonances, the absorption with the 3D photonic band gap back reflector is below the Lambertian limit. 
In contrast, at all six $s-$ and $p-$polarized resonances (S1-S4 and P1-P2) the 3D photonic band gap back reflector results exceed the Lambertian absorption, with the footnote that our calculation invokes a flat top surface and no optimized scattering surface. 
In the modern view, an absorption beyond Lambertian limit occurs if the local density of states (LDOS) inside the absorbing layer exceeds the LDOS outside~\cite{Saive2021ProgressinPV}. 
In one view point, the combination of ultrathin film with photonic back reflector leads to new resonances (S1-S4 and P1-P2, see below) that have a higher LDOS than the vacuum before the absorbing layer (above in Fig.~\ref{fig:SolarCellBackReflector}). 
Alternatively, since a photonic band gap blocks all incident radiation and inhibits the LDOS, a photonic band gap back reflector may be viewed as a peculiar ``colored electromagnetic vacuum''~\cite{Busch2000PRE} below the absorbing film that has a much lower LDOS than the one in the film itself. 
\begin{figure}[ht]
\centering
\includegraphics[width=0.65\columnwidth]{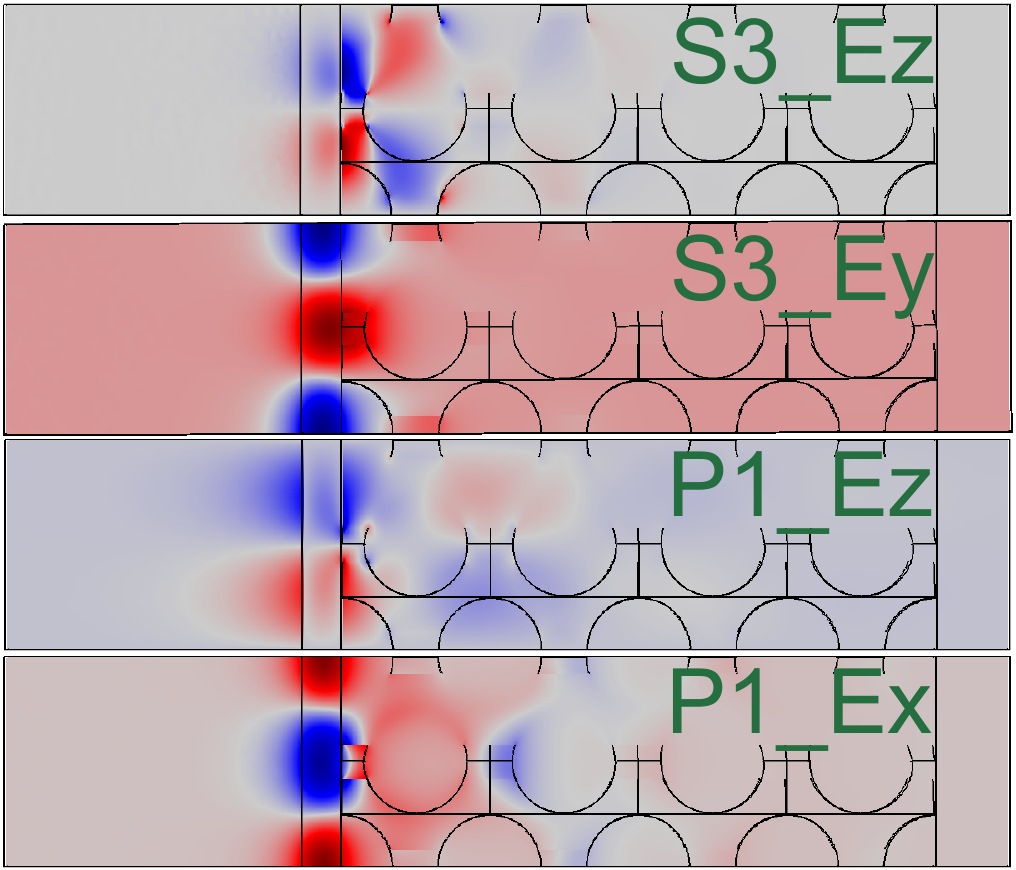}
\caption{Distribution of the electric field components for waves propagating in the sub-wavelength ultrathin film ($L_{Si}= 80$ nm) with a 3D photonic band gap back reflector. 
S3 and P1 are absorption resonances from Fig.~\ref{fig:SubwavelengthPS}, and we show field components that are not present in the incident waves. 
Red and blue denote the maxima and the minima of the electric field components, respectively.} 
\label{fig:EFieldSubwavelength}
\end{figure}

To investigate the physics behind these intriguing peaks, we discuss the electric field distributions for the exemplary resonances S3 and P1. 
The incident light with wavevector in the $\mathrm{Z}$-direction has either $s$ polarization (\textbf{E}-field in the $\mathrm{X}$-direction) or $p$ polarization (\textbf{E}-field in the $\mathrm{Y}$-direction). 
To filter the scattered electric fields that are guided in the plane of the ultrathin layer from possibly overwhelming incident fields, we plot in Fig.~\ref{fig:EFieldSubwavelength} the field components that are \textit{absent} in the incident light. 
Thus, we plot the ${E}_{z}$ and ${E}_{y}$ field components of the S3 resonance that is excited by $s-$polarized light ($\mathrm{E_{in,X}}$), and the ${E}_{z}$ and ${E}_{x}$ components of the P1 resonance that is excited by $p-$polarized light ($\mathrm{E_{in,Y}}$). 

First, we discuss the S3 ${E}_{y}$ and P1 ${E}_{x}$ field components that are transverse to the incident direction, and perpendicular to the incident polarization. 
Both components are maximal inside the ultrathin silicon film, as expected for guided waves. 
The field distributions show periodic variations in the plane of the ultrathin film that are also characteristic of guided modes~\cite{Fan2002PRB, Brongersma2014NatMat}. 
The periods match with the crystal's lattice parameter, which suggests that the fields are part of a Bloch mode. 
The fields extend into the photonic crystal by about one unit cell, in agreement with a Bragg attenuation length of about $\ell_{Br} = 0.6c = 180$ nm at the $p-$stop band center that was computed in Ref.~\cite{Devashish2017PRB}.

For the other two field components S3 ${E}_{z}$ and P1 ${E}_{z}$, the maximum fields are located at the air-silicon and at the silicon-photonic crystal interfaces, and the amplitudes decay away from the interfaces. 
The components show less conventional guiding behavior: the S3 ${E}_{z}$ component is maximal just \textit{inside} the photonic crystal (by about a quarter unit cell), and has a nodal plane parallel to the ultrathin film. 
The P1 ${E}_{z}$ component has maxima on either side of the ultrathin silicon film, somewhat like the field pattern of a long-range surface plasmon polariton (LRSPP) on a thin metal film~\cite{Berini2009AOP, DiffWithLRSPP}. 
This field extends about $\Delta \mathrm{Z} = 1$ unit cell into the crystal. 
These field distributions are the plausible signature of the confinement of a surface mode~\cite{Joannopoulos2008Book, Ishizaki2009NatLett}. 
Hence, the device can be viewed as an ultrathin absorbing dielectric film on top of a photonic crystal, which thus acts as a surface defect on the crystal and sustains a guided surface state. 
Consequently, the remarkable peaks observed in Fig.~\ref{fig:SubwavelengthPS} correspond to guided modes confined in an ultrathin layer consisting of two separate contributions: (i) a non-zero thickness layer due to the Bragg attenuation length of the crystal's band gap, and (ii) a deeply sub-wavelength ultrathin silicon film.    

\subsection{Wavelength-resolved transmission and absorption}
\label{subsec:wavelengthTransAbs}
\begin{figure}[ht]
\centering
\includegraphics[width=0.75\columnwidth]{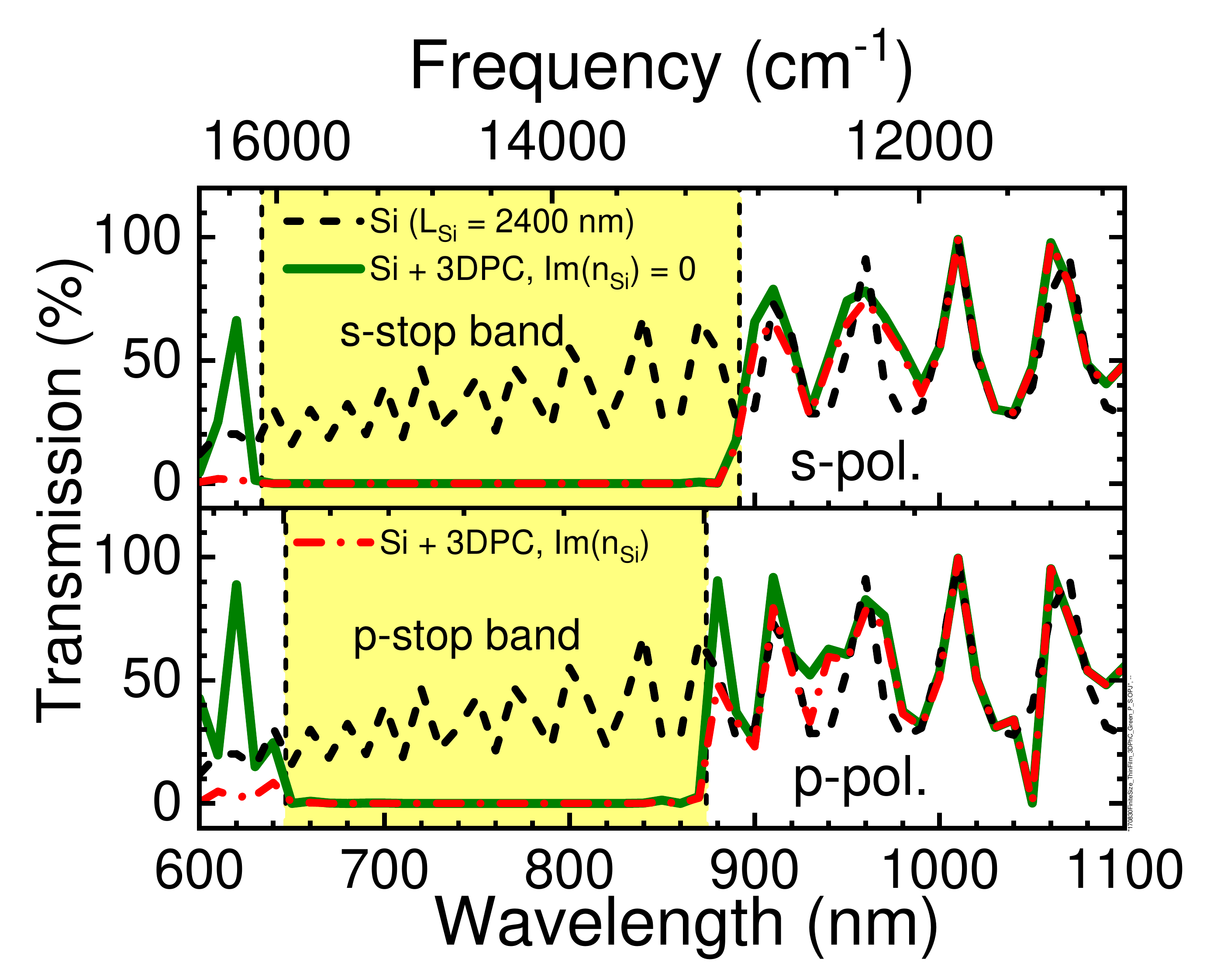}
\caption{Transmission spectra calculated for a thin silicon film ($L_{Si} = 2400$ nm) at normal incidence. 
Top panel: $s$ polarization, bottom panel: $p$ polarization. 
Black dashed curves are the transmission spectra for a thin film. 
Green solid curves are results for the thin film with a 3D inverse woodpile photonic crystal back reflector, with dispersion but no silicon absorption ($\mathbb{I}\mathrm{m}(n_{Si}) = 0$). 
Red dashed-dotted curves are results for the thin film with a 3D inverse woodpile photonic crystal back reflector, including silicon absorption ($\mathbb{I}\mathrm{m}(n_{Si}) \neq 0$). 
The vertical dashed lines are the edges of the $s-$ and $p-$stop bands that are shown as yellow bars.
} 
\label{fig:TransPS}
\end{figure}

To analyze the physics behind the results in Sections~\ref{subsec:ComparisonPerfectMetal} and~\ref{subsec:Subwavelength}, we break the problem down into several steps. 
Firstly, we briefly recapitulate the known situation of a thin silicon film only.  
Secondly, we study the thin film with a photonic crystal back reflector, where we only consider dispersion, but no absorption. 
This fictitious situation allows a comparison to the dispersion-free results that pertain to frequencies below the silicon band gap, see Ref.~\cite{Devashish2017PRB}. 
Thirdly, we study the complete device structure with the full silicon dispersion and absorption taken from Ref.~\cite{Green2008SolarEnergyMater}. 

Figure~\ref{fig:TransPS} reveals oscillations between $\lambda = 600$ nm and $\lambda = 1100$ nm for both polarizations in the transmission spectra for the thin silicon film ($L_{Si} = 2400$ nm). 
These oscillations are Fabry-P{\'e}rot fringes~\cite{Yariv1984Book} resulting from multiple reflections of the waves inside the film at the front and back surfaces of the thin silicon film. 

In Fig.~\ref{fig:TransPS}, we observe nearly $0\%$ transmission between $\lambda = 600$ nm and $\lambda = 900$ nm for a thin film with a photonic crystal back reflector, both with and without silicon absorption. 
These deep transmission troughs are photonic stop bands between $\lambda = 647$ nm to $874$ nm for $p$ polarization and $\lambda = 634$ nm to $892$ nm for $s$ polarization. 
These gaps were previously identified to be the dominant stop gaps in the $\Gamma-\mathrm{X}$ and $\Gamma-\mathrm{Z}$ high-symmetry directions (see Appendix~\ref{sect:Brillouinzone}) that encompass the 3D photonic band gap~\cite{Devashish2017PRB}. 
From the good agreement of the stop bands, both with and without silicon absorption, we deduce that an inverse woodpile photonic crystal behaves as a perfect reflector in the visible range, even in presence of substantial absorption. 
This result further supports the above observation that the absorption length of silicon is much longer than the Bragg attenuation length of the 3D inverse woodpile photonic crystal $l_{a} >> \ell_{Br}$. 
Hence, waves incident on the photonic crystal are reflected before being absorbed by the high-index backbone of the photonic crystal. 
Thus, the Fabry-P{\'e}rot fringes in transmission between $\lambda = 600$ nm and $\lambda = 860$ nm are completely suppressed by strong and broadband reflection of the 3D photonic crystal back reflector~\cite{Devashish2017PRB}. 

\begin{figure}[ht]
\centering
\includegraphics[width=0.85\columnwidth]{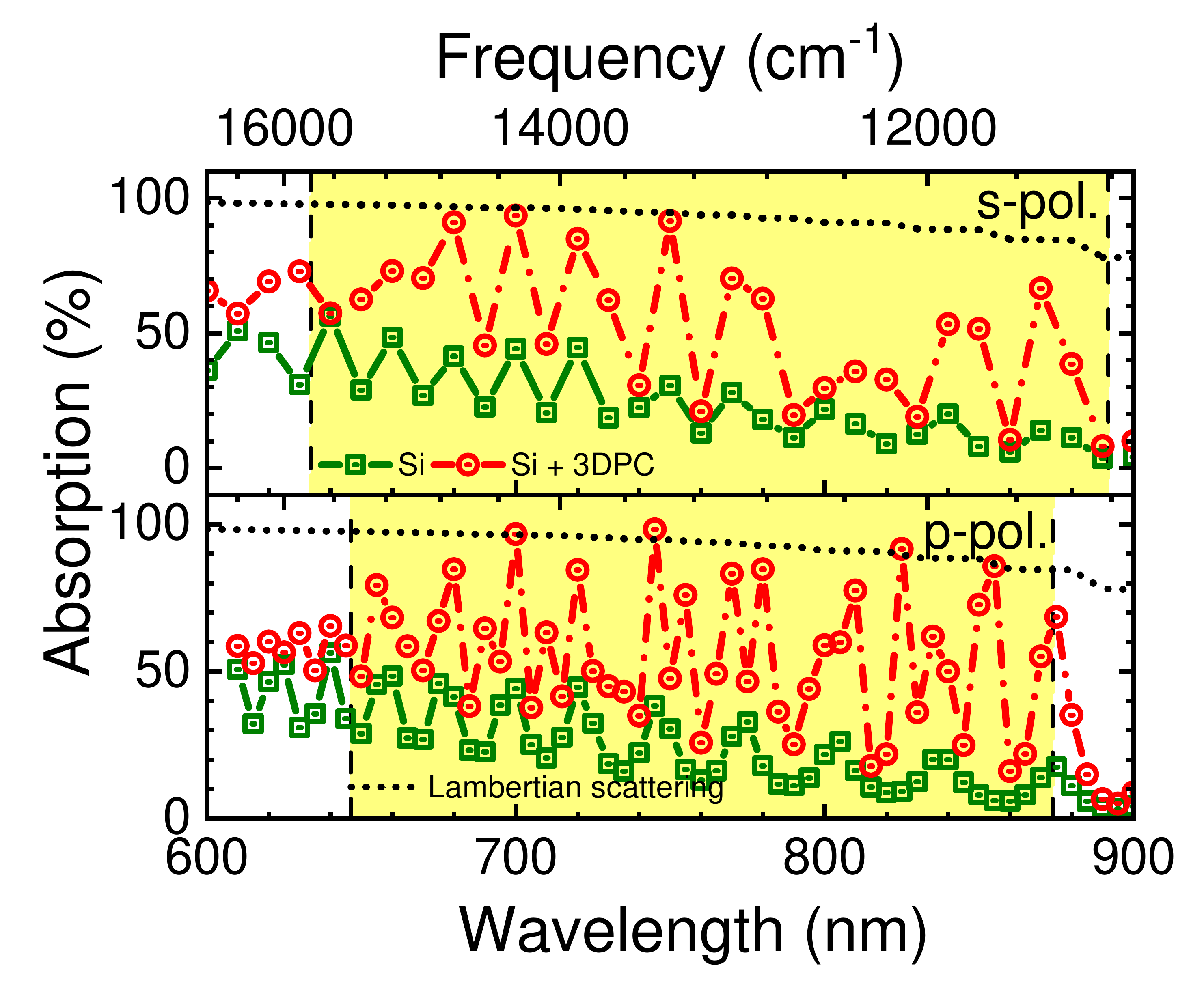}
\caption{
Absolute absorption (in $\%$) by a thin silicon film ($L_{Si} = 2400$ nm) in the stop band of a 3D inverse woodpile photonic crystal. 
Top panel: $s-$polarized, bottom panel: $p-$polarized light. 
Green solid curves are absorption spectra for a thin silicon film. 
Red dashed-dotted curves are absorption spectra for a thin silicon film with a 3D inverse woodpile photonic crystal back reflector. 
The vertical dashed lines are the edges of the $s-$ and $p-$stop bands.
The stop bandwidths are shown as the yellow bar. 
Black dotted curves represent absorption $A^{L}_{Si}(\lambda)$ with Lambertian scattering, from Eq.~\ref{eq:LambertianAbsorption}.
} 
\label{fig:ZoomedAbsPS}
\end{figure}

Furthermore, oscillations below $\lambda = 600$ nm are present in the transmission spectra for the thin silicon film with a photonic crystal back reflector without absorption, but not in the system with absorption. 
Since the transmission spectra with realistic absorption show nearly $0\%$ transmission below $\lambda = 600$ nm, where silicon is strongly absorbing, all light is absorbed and the Fabry-P{\'e}rot fringes are suppressed. 

Figure~\ref{fig:ZoomedAbsPS} shows the absorption spectra for a thin silicon film without and with a 3D photonic crystal back reflector. 
We consider the dispersive and complex refractive index for the silicon and the high-index backbone of the photonic crystal. 
Fabry-P{\'e}rot fringes appear below $\lambda = 900$ nm, corresponding to standing waves in the thin silicon film. 
Since the imaginary part of the silicon refractive index increases with decreasing wavelength (see Fig.~\ref{fig:nSi}), the absorption in silicon also increases with decreasing wavelength. 

Between $\lambda = 600$ nm and $\lambda = 900$ nm in the top and bottom panels of Fig.~\ref{fig:ZoomedAbsPS}, there are more Fabry-P{\'e}rot fringes. 
The fringes have a greater amplitude for a thin silicon film with photonic crystal back reflector, compared to a standalone thin silicon film. 
To interpret the increased number of fringes, we consider diffraction from the photonic crystal surface. 
In Fig.~\ref{fig:SolarCellBackReflector}, light first travels through silicon before reaching the photonic crystal surface. 
Therefore, the incident wavelength reduces to $\lambda / n_{Si}$ inside the silicon layer. 
In the entire stop bands (both $s$ and $p$) the wavelength $\lambda / n_{Si}(\lambda)$ between $\lambda = 600$ nm and $\lambda = 900$ nm is smaller than the lattice parameter $c = 300$ nm of the 3D inverse woodpile photonic crystal (along the $\Gamma \mathrm{Z}$ direction). In addition, a 3D inverse woodpile photonic crystal introduces a periodic refractive index contrast at the interface with a thin silicon film. 
Hence, nonzero diffraction modes~\cite{Joannopoulos2008Book, Biswas2010SolarEnergyMatSolarCells} are generated at the photonic crystal-thin silicon film interface at specific wavelengths inside the stop bands, resulting in additional reflected waves that are then absorbed in the film.

Between $\lambda = 600$ nm and $900$ nm, the silicon thickness $L_{Si} = 2400$ nm is larger than half the wavelength $\lambda / n_{Si}$ in the absorbing layer. 
This is the condition for a photonic crystal-thin silicon film interface to couple the reflected waves into the guided modes~\cite{Griffiths1998Book, OBrien2008AdvMater} that propagate inside the silicon. 
Hence, the physical mechanism responsible for the additional number of fringes is the occurrence of non-zero diffraction modes coupled into guided modes due to the photonic crystal back reflector. 
The absorption in guided modes can sometimes approach $100\%$~\cite{OBrien2008AdvMater}, \textit{e.g.}, at $\lambda = 700$ nm and $720$ nm in Fig.~\ref{fig:ZoomedAbsPS} (bottom). 
We note that the perfect reflectivity of a 3D inverse woodpile photonic crystal extends over the entire stop bands, whereas nonzero order diffraction modes and guided modes are limited to specific wavelengths. 
This is in contrast to Section~\ref{subsec:Subwavelength}, where we studied a reduced thickness such that no guided modes are allowed. 

For reference, Fig.~\ref{fig:ZoomedAbsPS} also shows the absorption limit due to Lambertian scattering (Eq.~\ref{eq:LambertianAbsorption}). 
In most of the spectral range, the absorption with the 3D photonic band gap back reflector is below the Lambertian limit. 
At several resonances, the photonic band gap back reflector results match or even slightly exceed the Lambertian limit, again with the note that our calculation does not invoke an optimized scattering surface. 
The fact that at this thickness the results exceed the Lambertian limit less than in the earlier ultrathin case makes intuitive sense; if we consider the silicon film as a Fabry-P{\'e}rot resonator, and we consider a certain constant wavelength, then a thicker resonator corresponds to a greater reduced frequency ($L_{Si} / \lambda$) and will show less LDOS modulation than a thin low ($L_{Si} / \lambda$) resonator. 
Therefore~\cite{Saive2021ProgressinPV}, the thick resonator is less likely to reveal beyond-Lambertian absorption, in agreement with our results. 

\subsection{Negligible absorption inside the photonic crystal backbone}\label{subsect:backbone}
Since we propose the back reflector to be a photonic crystal that is also made of silicon, one might rightfully hypothesize that the mere presence of extra material in the photonic crystal simply increases the total length of silicon, which thus \textit{sneakily} enhances the absorption. 
To evaluate this hypothesis, we compute and compare the absorption for three different devices. 
We first study a thin silicon film of thickness $L_{Si} = 2400$ nm without back reflector, neither perfect metal nor photonic band gap (device $\# 1$).
Secondly, we consider a silicon film ($L_{Si} = 2400$ nm) with a photonic crystal back reflector with a thickness $L_{Si} = 1200$ nm; this device $\# 2$ has a total thickness of $3600$~nm. 
Thirdly, we study a structure with the same total thickness as the second one, namely a silicon layer with a thickness $L_{Si} = 3600$ nm, but without back reflector (device $\# 3$).  

\begin{figure}[ht]
\centering
\includegraphics[width=0.75\columnwidth]{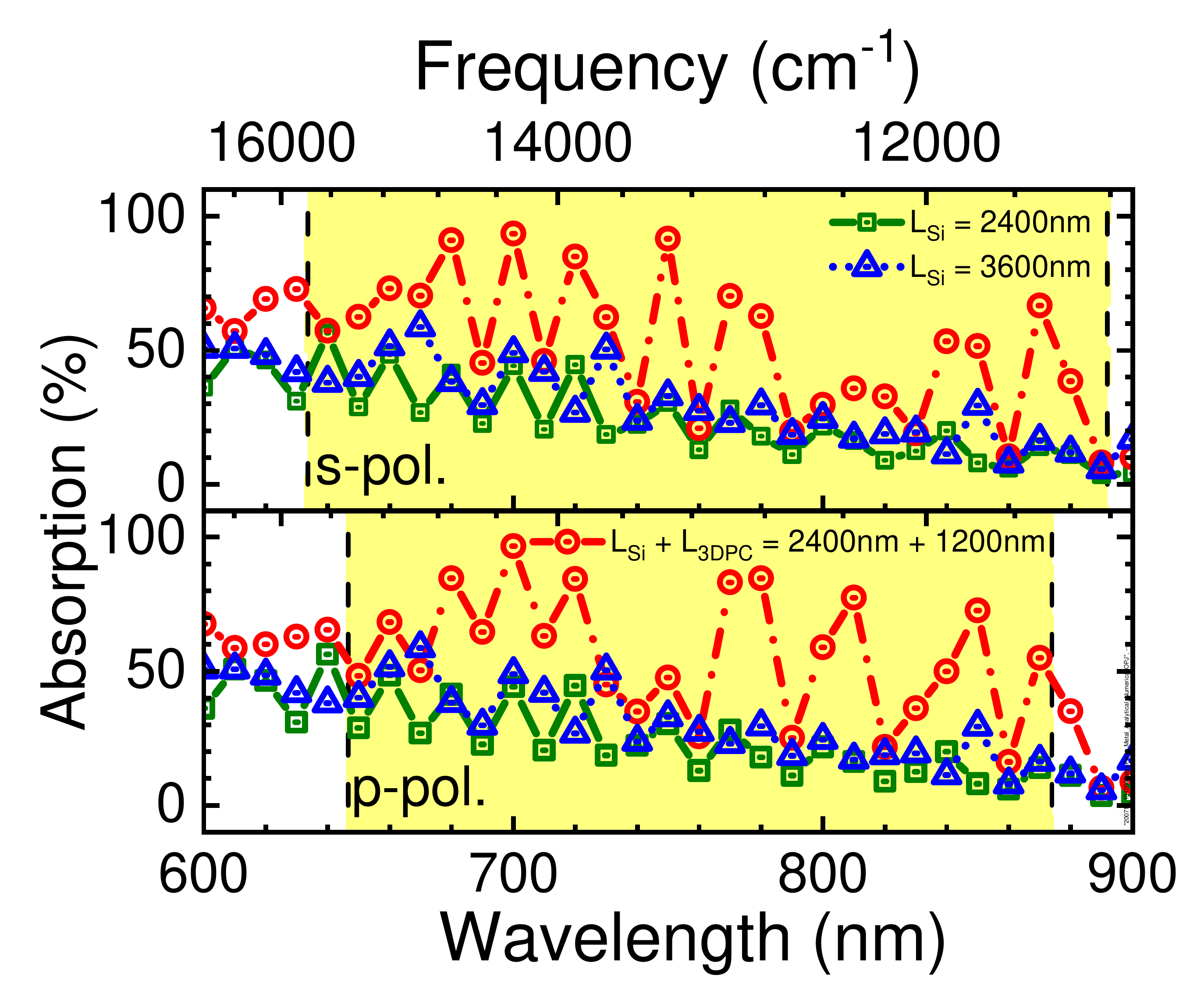}
\caption{Verification of the absence of absorption inside the photonic crystal backbone. 
Top panel: $s-$polarized, bottom panel: $p-$polarized light, with $s-$ and $p-$stop bands shown as yellow highlighted regions. 
Green solid curves pertain to a thin film only with the silicon thickness $L_{Si} = 2400$ nm.
Red dashed-dotted curves pertain to a thin silicon film (thickness $L_{Si} = 2400$ nm) with a 3D photonic crystal back reflector (thickness $L_{3DPC} = 1200$ nm). 
Blue dotted curves pertain to a thin film only with the same overall thickness $L_{Si} = 3600$ nm. 
} 
\label{fig:AbsPS8c12c}
\end{figure}

To investigate the effect of the band gap on the absorption in the thin film, we zoom in on the absorption spectra inside the stop bands in Fig.~\ref{fig:AbsPS8c12c}. 
We observe that the absorption spectra are closely the same for both thin films (devices $\# 1$ and $\# 3$), including the Fabry-P{\'e}rot fringes. 
Since film $\# 3$ is considerably thicker than $\# 1$, the similarity strongly suggests that the silicon absorption within the stop band wavelength range is saturated for the thinner layer. 

In contrast, device $\# 2$ with a photonic crystal back reflector reveals significantly higher absorption, including a larger amplitude of the Fabry-P{\'e}rot fringes than the films $\# 1$ and $\# 3$. 
Since the total thickness of device $\# 2$ is the same as for film $\# 3$, we conclude that the enhanced absorption of the film with the photonic band gap back reflector is \textit{not} caused by the additional thickness of the back reflector itself, hence the \textit{sneaky} effect does not exist. 
The conclusion that the back reflector does not absorb is also reasonable, since we have seen above that the absorption of the light mostly occurs within the films themselves. 
Using Eq.~\ref{eq:EnhancementRatio}, we find that the wavelength-averaged absorption enhancement due to a photonic crystal back reflector is nearly $\langle\eta^{'}_{abs}\rangle = 1.8 \times$ for the $s-$stop band and nearly $\langle\eta^{'}_{abs}\rangle = 1.9 \times$ for the $p-$stop band compared to a thin film with thickness $L_{Si} = 3600$ nm. 

\section{Practical considerations for devices}\label{sec:practical}

\subsection{Angular acceptance}\label{sec:angular}

\begin{figure}[ht]
    \centering
    \begin{minipage}[b]{0.47\textwidth}
        \centering
        \includegraphics[width=\textwidth]{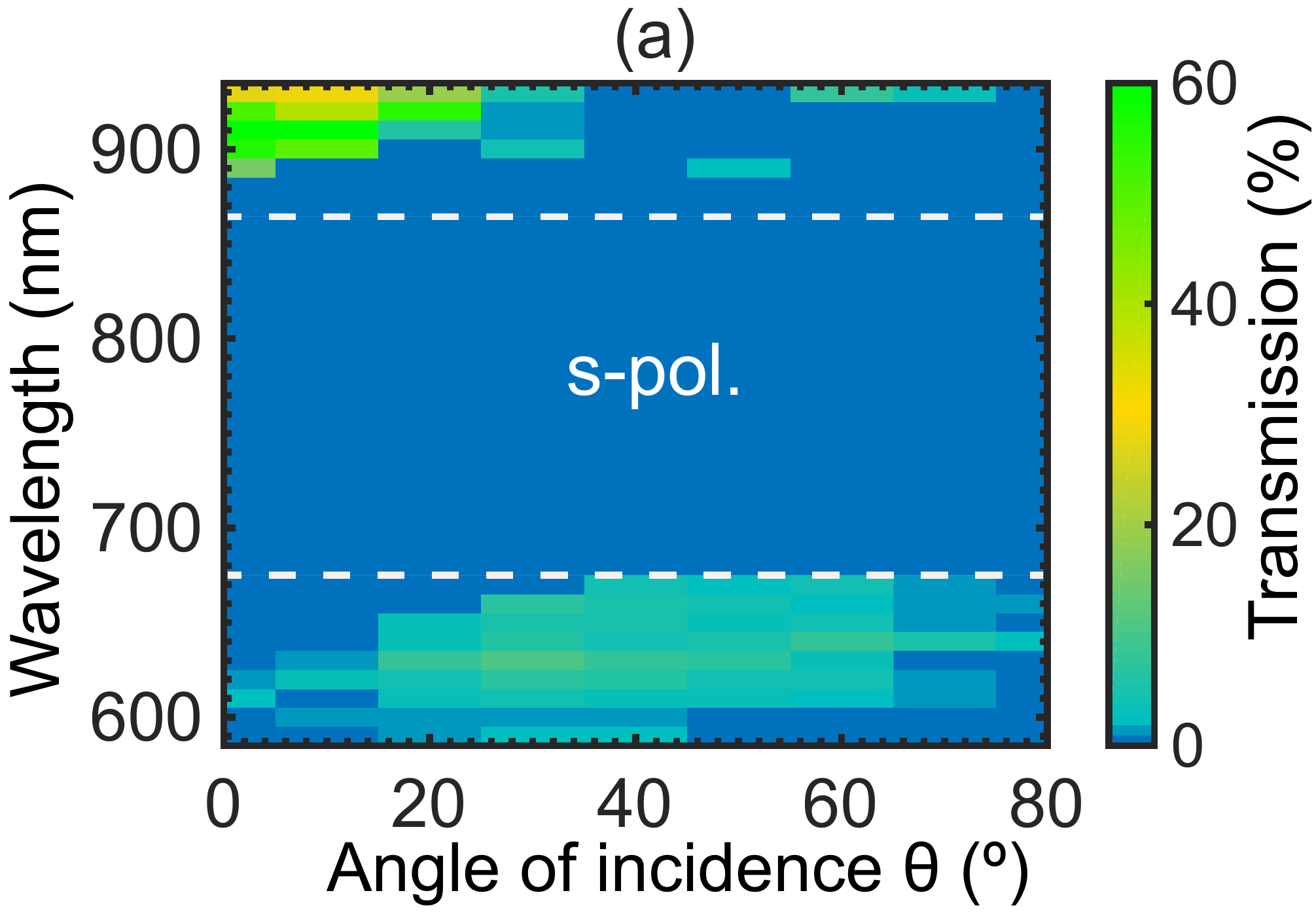}   
        \label{fig:AngleFreqTransS}
    \end{minipage}
    \begin{minipage}[b]{0.49\textwidth}  
        \centering 
        \includegraphics[width=\textwidth]{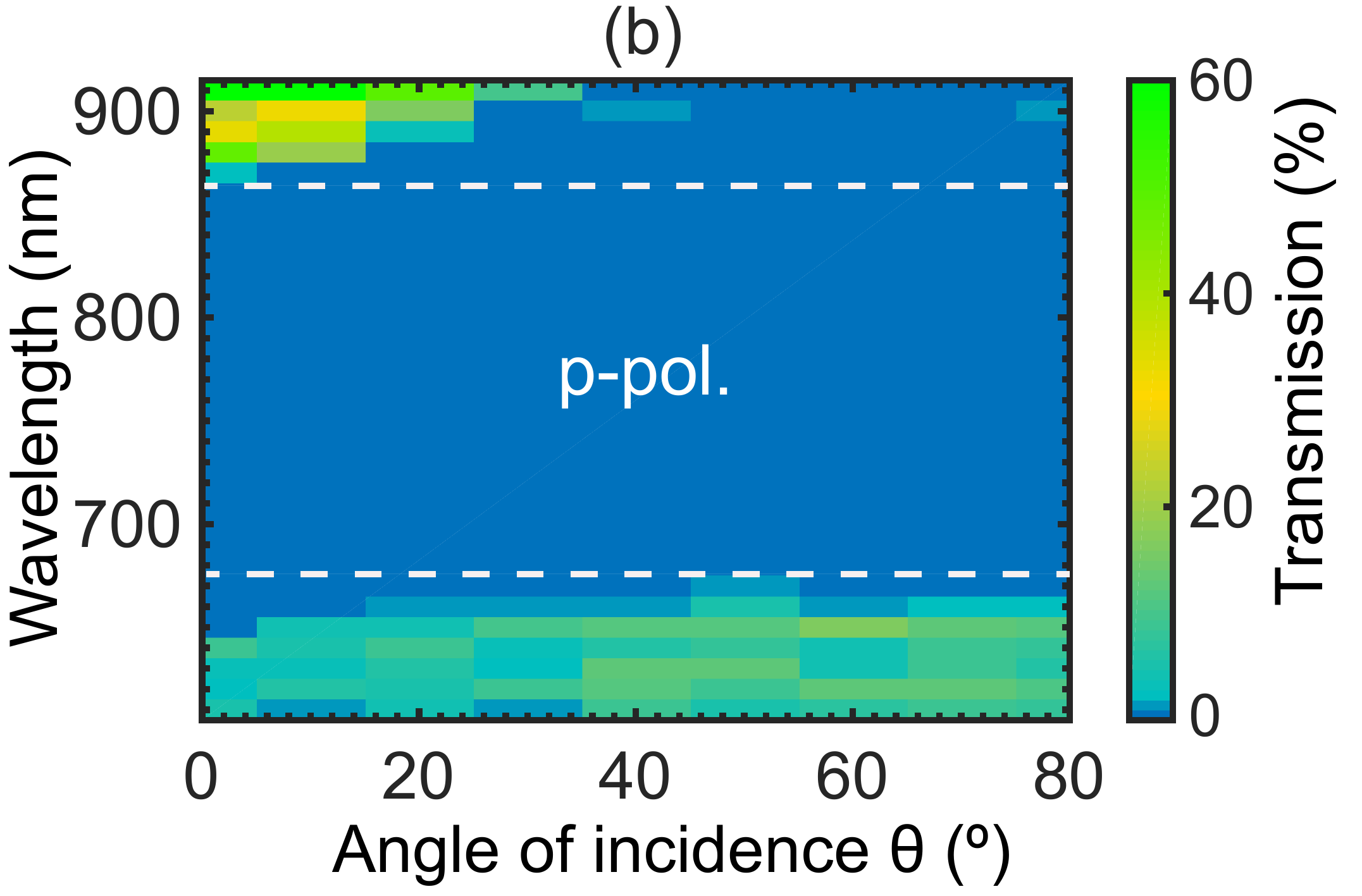}
        \label{fig:AngleFreqTransP}
    \end{minipage}
    \caption{Angle- and wavelength-resolved transmission spectra calculated for a thin silicon film ($L_{Si} = 2400$ nm) with a 3D inverse woodpile photonic crystal back reflector for (a) $s$ polarization and (b) $p$ polarization. 
    The dark blue color represents nearly $0\%$ transmission that occurs in the stop band at all incident angles. 
    The white dashed box indicates an angle- and polarization-independent range with nearly $0\%$ transmission.}
    \label{fig:AngleFreqTransPS}
\end{figure}

In order to optimize the absorption of a thin silicon film, we first investigate the impact of a 3D photonic crystal back reflector on the angular acceptance. 
Figures~\ref{fig:AngleFreqTransPS} (a, b) show transmission maps versus angle of incidence and wavelength. 
The angle of incidence is varied up to $\theta = 80^{\circ}$ off the normal.  
For both polarizations \textit{simultaneously} we observe a broad angle-independent stop band that is characterized by near $0\%$ transmission. 
The broad stop band extends all the way from $\lambda = 680$ nm to $\lambda = 880$ nm. 
This shows that the Bragg attenuation length for the 3D inverse woodpile photonic crystal is smaller than the absorption length of silicon for all incident angles for the omnidirectional stop band. 
Thus, a 3D photonic band gap crystal acts as a perfect reflector in the omnidirectional stop band for all incident angles and for all polarizations, even with full absorption in the refractive index 
- in other words, an \textit{omnidirectional} stop band.

\begin{figure}[]
\centering
\includegraphics[width=0.75\columnwidth]{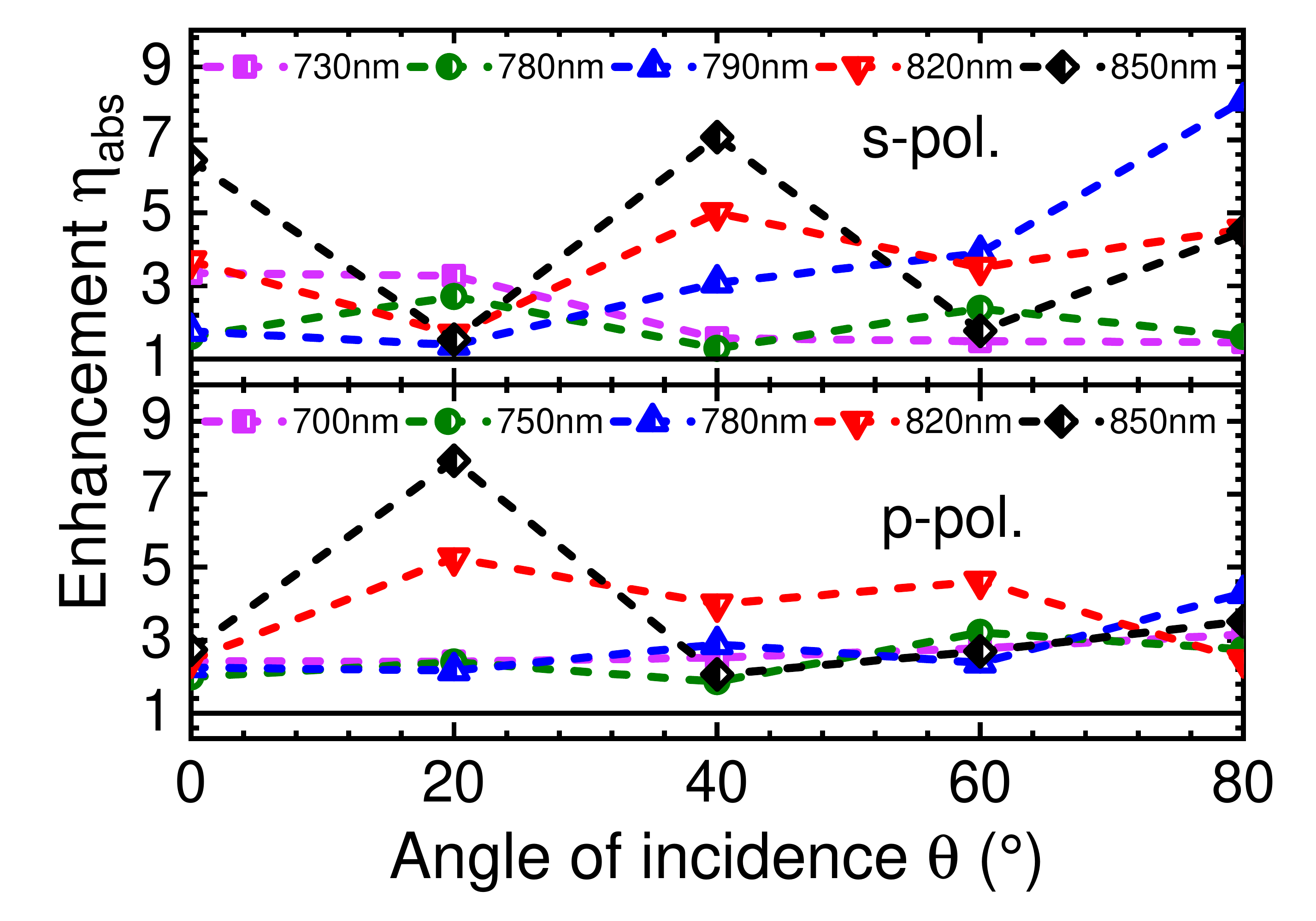}
\caption{Absorption enhancement $\eta_{abs}(\lambda, \theta)$ versus incidence angle $\theta$ off the surface normal to characterize the angular acceptance of a thin silicon film ($L_{Si} = 2400$ nm) with a 3D inverse woodpile back reflector, for 5 wavelengths throughout the band gap (850 nm, 820 nm, 790 nm, 760 nm, and 730 nm) with connected symbols and colors as shown in the legend.
The top panel is for $s-$polarized light, and the bottom panel for $p-$polarized light. 
} 
\label{fig:AngleEnhancementPS}
\end{figure}

Figure~\ref{fig:AngleEnhancementPS} shows the absorption enhancement $\eta_{abs}(\lambda, \theta)$ versus incident angle for five representative wavelengths throughout the band gap of a 3D inverse woodpile back reflector. 
For all incident angles up to $80^{\circ}$, we observe that the absorption enhancement stays above 1, which corresponds a standalone thin film. 
Furthermore, at certain incident angles, the absorption enhancement is as high as 7 or 9. 
For both polarizations, Fig.~\ref{fig:AngleEnhancementPS} also reveals oscillatory absorption enhancement with increasing incident angles, which are signature of the Fabry-P{\'e}rot fringes~\cite{Yariv1984Book}. 
Hence, a 3D inverse woodpile crystal widens the angular acceptance of a thin silicon film by creating an omnidirectional absorption enhancement regime.

To calculate the angle-averaged and wavelength-averaged absorption enhancement $\langle\eta^{''}_{abs}\rangle$ in the \textit{omnidirectional stop band}, we employ Eq.~\ref{eq:AllAnglesEnhancementRatio}.
Consequently, the angle- and wavelength-averaged absorption enhancement for $s$ polarization is $\langle\eta^{''}_{abs}\rangle$ $= 2.11\times$ and for $p$ polarization is $\langle\eta^{''}_{abs}\rangle$ $= 2.68\times$, which exceeds the maximum absorption enhancement feasible for a perfect metallic reflector. 
These enhancements are possible only if a photonic crystal back reflector generates non-zeroth order diffraction modes at certain discrete wavelengths for all incident angles~\cite{Yu2011APL}. 
Once these non-zero diffraction modes couple into guided modes and are confined inside the thin film via total internal reflection, the effective optical path length travelled is longer than the one travelled by a zero order diffraction mode. 
Therefore, a 3D inverse woodpile crystal enhances the absorption of a thin silicon film for all incident angles and polarizations by (i) revealing perfect reflectivity inside the omnidirectional stop band and (ii) generating guided modes for specific wavelengths.


\subsection{Optimal thickness of the absorbing thin film}\label{sec:thickness} 

\begin{figure}[ht]
\centering
\includegraphics[width=0.75\columnwidth]{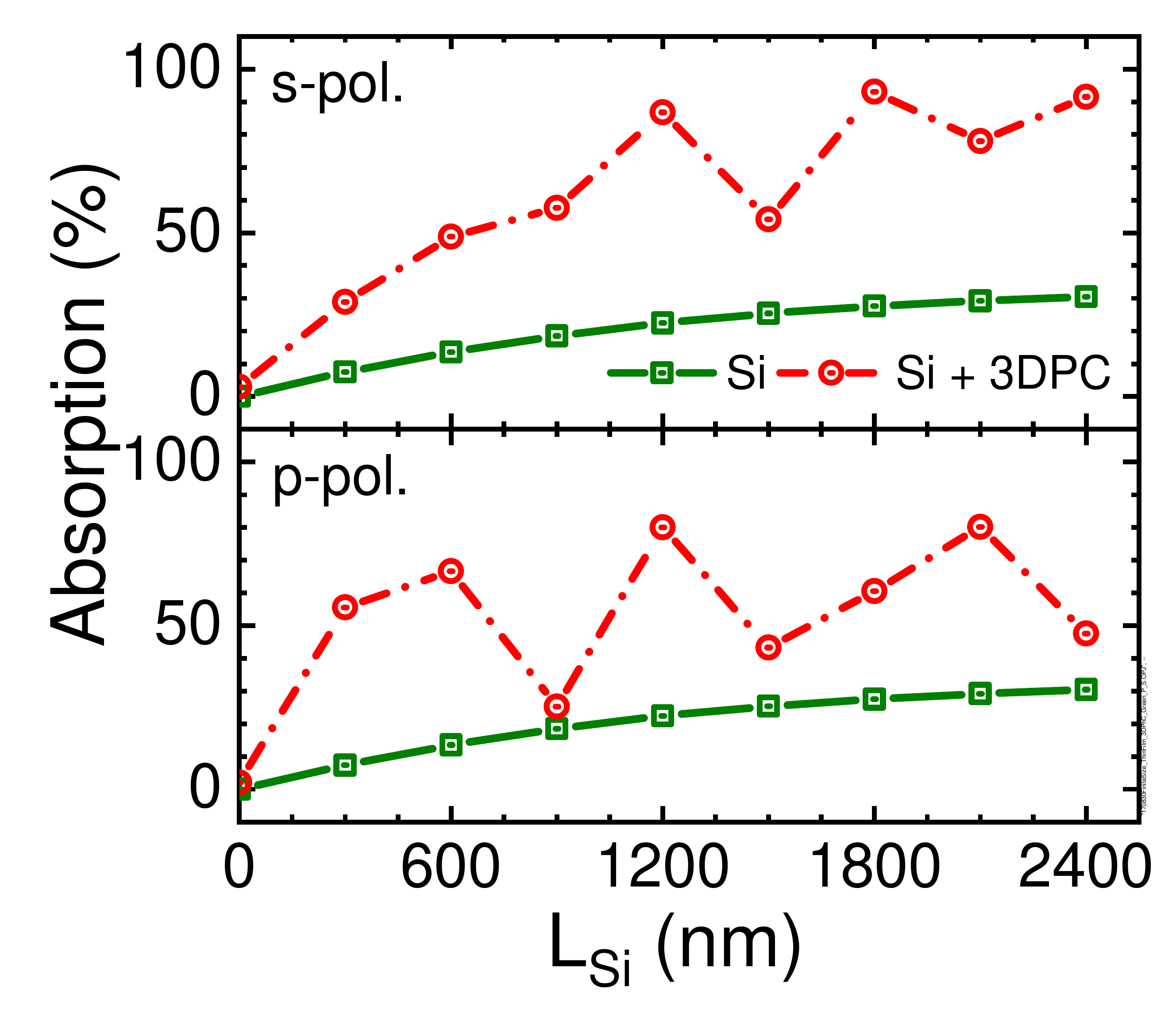}
\caption{Absorption versus thickness $L_{Si}$ of a thin silicon film in presence (connected red circles) of a 3D photonic crystal back reflector, and without back reflector (green squares). 
The data pertain to a wavelength $\lambda = 760$ nm, the 3D photonic band gap center of the back reflector.
Top panel: $s-$polarized, bottom panel: $p-$polarized light.
} 
\label{fig:FiniteSizePS}
\end{figure}

To investigate the effect of the thickness $L_{Si}$ of the thin silicon film on its absorption, we plot in  Fig.~\ref{fig:FiniteSizePS} the absorption for light at normal incidence for film thicknesses between $L_{Si} = 300$ nm and $L_{Si} = 2400$ nm in presence of a 3D inverse woodpile photonic crystal back reflector with a constant thickness $L_{3DPC} = 1200$ nm. 
While the absorption for a thin film shows a monotonic increase with increasing silicon thickness for both polarizations, Fig.~\ref{fig:FiniteSizePS} reveals that the absorption for a thin film with a photonic crystal back reflector is always higher than the corresponding standalone thin film. 
Furthermore, absorption spectra of a thin film with a photonic crystal back reflector reveals oscillations with increasing film thicknesses for both polarizations, showing saturation towards higher wavelength. 
We surmise that in order to maximize absorption enhancement for a given wavelength, the thickness of a thin silicon film had better be chosen to the maxima of the oscillations in Fig.~\ref{fig:FiniteSizePS}. 
Therefore based on this normal incidence analysis, when designing a device the thickness of the silicon film had better be tweaked from $L_{Si} = 2400$ nm to one of the fringe maxima in Fig.~\ref{fig:FiniteSizePS}, such as $L_{Si} = 1200$ nm or the range $1800-2100$ nm. 

\subsection{Other experimental considerations}\label{sec:considerations} 
To enhance the absorption of light over an even broader wavelength range than reported here, it is relevant to consider different orientations of the 3D photonic crystal back reflector. 
In case of both direct and inverse woodpile structures, it is interesting to consider light incident in the $\Gamma \mathrm{Y}$ direction, since the $\Gamma \mathrm{Y}$ stop gap with a relative bandwidth $39.1~\%$ (see Fig.~2 of Ref.~\cite{Devashish2017PRB}) is about $1.3 \times$ broader than the $\Gamma \mathrm{Z}$ or $\Gamma \mathrm{X}$ stop gaps whose relative bandwidth is $30.4~\%$~\cite{Hillebrand2003JAP, Woldering2009JAP, Huisman2011PRB, Devashish2017PRB}. 

To improve the likelihood that photonic band gap back reflectors gain real traction, it is obviously relevant to consider strategies whereby such a back reflector can be realized over as large as possible ($\mathrm{X}$,$\mathrm{Y}$) areas. 
In the current nanofabrication of silicon inverse woodpiles, an important limitation is the depth of the nanopores that are fabricated by deep reactive-ion etching~\cite{vandenBroek2012AFM, Grishina2019ACS}, which limits the $\mathrm{X}$- or $\mathrm{Z}$-extent of the nanostructures, whereas the $\mathrm{Y}$-extent has no fundamental limit. 
Therefore, with the design shown in Fig.~\ref{fig:ComputationalCell}, the back reflector would have sufficient $\mathrm{Z}$-extent (thickness) and a large $\mathrm{Y}$-extent, but limited $\mathrm{X}$-extent, and thus limited area. 
A remedy consists of etching the nanopores at $45^\circ$ to the back surface, as demonstrated by Takahashi \textit{et al.}~\cite{Takahashi2006APL}. 
Then, the areal ($\mathrm{X}$,$\mathrm{Y}$)-extent of the photonic crystal equals the areal extent of the pore array, 
which can be defined by standard optical lithography or by self-assembly. 
In such a design, light at normal incidence to the thin silicon film arrives at $45^\circ$ with respect to the inverse woodpile structure, parallel to the $\Gamma \mathrm{U}$ high-symmetry direction. 
The stop gap for this high-symmetry direction has properties that are fairly similar to the $\Gamma \mathrm{Z}$ or $\Gamma \mathrm{X}$ stop gaps considered here~\cite{Hillebrand2003JAP, Woldering2009JAP, Huisman2011PRB, Devashish2017PRB}. 
Therefore, the present analysis also pertains to this design. 

Having a 3D silicon photonic crystal as a back reflector provides an all-silicon integration with the absorbing thin silicon film. 
Moreover, this approach makes the thin film device lighter, since the 3D inverse woodpile photonic crystal structure is highly porous, consisting of nearly $80\%$ volume fraction air (see Appendix~\ref{sect:VolumeFractionInverseWoodpile}). 
Simultaneously, our design remarkably enhances the overall absorption in comparison to a standalone thin silicon film with the same overall thickness, as illustrated by our results. 

If one wishes to shift the absorption enhancements discussed here to shorter wavelengths, it is relevant to consider replacing silicon by wider band gap semiconductors such as GaAs, GaP, or GaN. 
In such a case, an important practical consideration is whether to realize the photonic band gap back reflector from the same semiconductor for convenient integration, or whether to perform heterogeneous integration of silicon photonic band gap crystals, since the latter are readily realizable, see Refs.~\cite{Tjerkstra2011JVSTB, vandenBroek2012AFM, Grishina2015NT, Grishina2019ACS}. 

Since GaAs has a similar real refractive index as silicon, many of the results presented here can be exploited to make predictions for GaAs absorption in presence of a photonic band gap back reflector. 
For the other semiconductors, this calls for additional detailed calculations, since most semiconductors have different (complex) refractive indices than silicon, with different dispersion. 
Nevertheless, the computational concepts and strategies presented in our study remain relevant to address the pertinent questions. 

To obtain an interpretation of Fig.~\ref{fig:EnhancementPS} and Fig.~\ref{fig:SubwavelengthPS} for photovoltaics where incident light is unpolarized, one can take the mean absorption enhancements $\eta_{abs}^{'}(\lambda)$ for $s$ polarization and $p$ polarization. 
As mentioned earlier, for the sake of physical understanding our computations pertain to a flat top surface of the thin silicon film without anti-reflection coating. 
On the other hand, it is well known that a substantial improvement in light harvesting is obtained by applying anti-reflection coatings and by tailoring the shape of top surface, for instance, a random Lambertian-like surface. 
Therefore, a logical research step is to combine a photonic band gap back reflector with suitable top-surface engineering, even though it is likely that this situation substantially complicates both the numerical setup and the numerical convergence.

\section{Conclusion}\label{sec:conclusions}
We investigated a thin 3D photonic band gap crystal as a back reflector in the visible regime, which reflects light within the band gap for all directions and for all polarizations. 
The absorption spectra of a thin silicon film with a 3D inverse woodpile photonic crystal back reflector were calculated using finite-element computations of the 3D time-harmonic Maxwell equations. 
We tailored the finite-sized inverse woodpile crystal design to have a broad photonic band gap in the visible range and have used the refractive index of the real silicon, including dispersion and absorption, in order to make our calculations relevant to experiments. 
From the comparison of the photonic crystal back reflector to a perfect metallic back reflector, we infer that a photonic crystal back reflector increases the number of Fabry-P{\'e}rot fringes for a thin silicon film. 
Therefore, we observe that a 3D inverse woodpile photonic crystal enhances the absorption of a thin silicon film by (i) behaving as a perfect reflector, exhibiting nearly 100\% reflectivity in the stop bands, as well as (ii) generating guided modes at many discrete wavelengths. 
Our absorption results show nearly $2.39 \times$ enhanced wavelength-, angle-, polarization-averaged absorption between $\lambda = 680$ nm and $\lambda = 880$ nm compared to a 2400 nm thin silicon film. 
We find that the absorption enhancement is enhanced by positioning an inverse woodpile back reflector at the back end of a thin silicon film, which will keep the length of the thin film device unchanged as well as make it lighter. 
In order to maximize the efficiency for a given wavelength, we show that the thickness of a thin silicon film had better be chosen to the maxima of the Fabry-P{\'e}rot fringes. 
For a sub-wavelength ultrathin 80 nm absorbing layer with a photonic crystal back reflector, we identify and demonstrate two physical mechanisms causing the giant average absorption enhancement of 9.15$\times$ : (i) guided modes due to the Bragg attenuation length and (ii) confinement due to a surface defect. 

\section{Funding}
\label{sect:Funding}
This research is supported by the Shell-NWO/FOM programme ``Computational Sciences for Energy Research" (CSER), NWO-FOM program nr. 138 ``Stirring of light!," NWO-TTW Perspectief program P15-36 ``Free-form scattering optics" (FFSO), NWO ENW-GROOT program (OCENW.GROOT.2019.071) "Self-assembled icosahedral photonic quasicrystals with a band gap for visible light", the ``Descartes-Huygens" prize of the French Academy of Sciences (with support from JMG), and the MESA$^{+}$ Institute section Applied Nanophotonics (ANP). 

\section{Acknowledgments}
\label{sect:Acknowledgements}
It is a great pleasure to thank Bill Barnes (Exeter and Twente), Ad Lagendijk, Allard Mosk (Utrecht University), Oluwafemi Ojambati, Pepijn Pinkse, and Ravitej Uppu (now at University of Iowa) for stimulating discussions, Diana Grishina for logistic support, and Femius Koenderink (now at AMOLF, Amsterdam) for the analytical expression of the volume fraction in the early days when we had just started developing silicon inverse woodpile photonic band gap crystals. We thank two anonymous reviewers for their very helpful suggestions. 

\section{Disclosures}
\label{sect:Disclosures}
The authors declare no conflicts of interest.

\section{Data availability}
\label{sect:DataAvailability}
Data underlying the results presented in this paper are not publicly available at this time but may
be obtained from the authors upon reasonable request. 
\appendix

\section{Primitive unit cell of the 3D inverse woodpile photonic crystal structure}
\label{sect:VolumeFractionInverseWoodpile}

 \begin{figure}[ht]
    \centering
        \begin{minipage}[b]{0.5\textwidth}   
            \centering 
            \includegraphics[width=\textwidth]{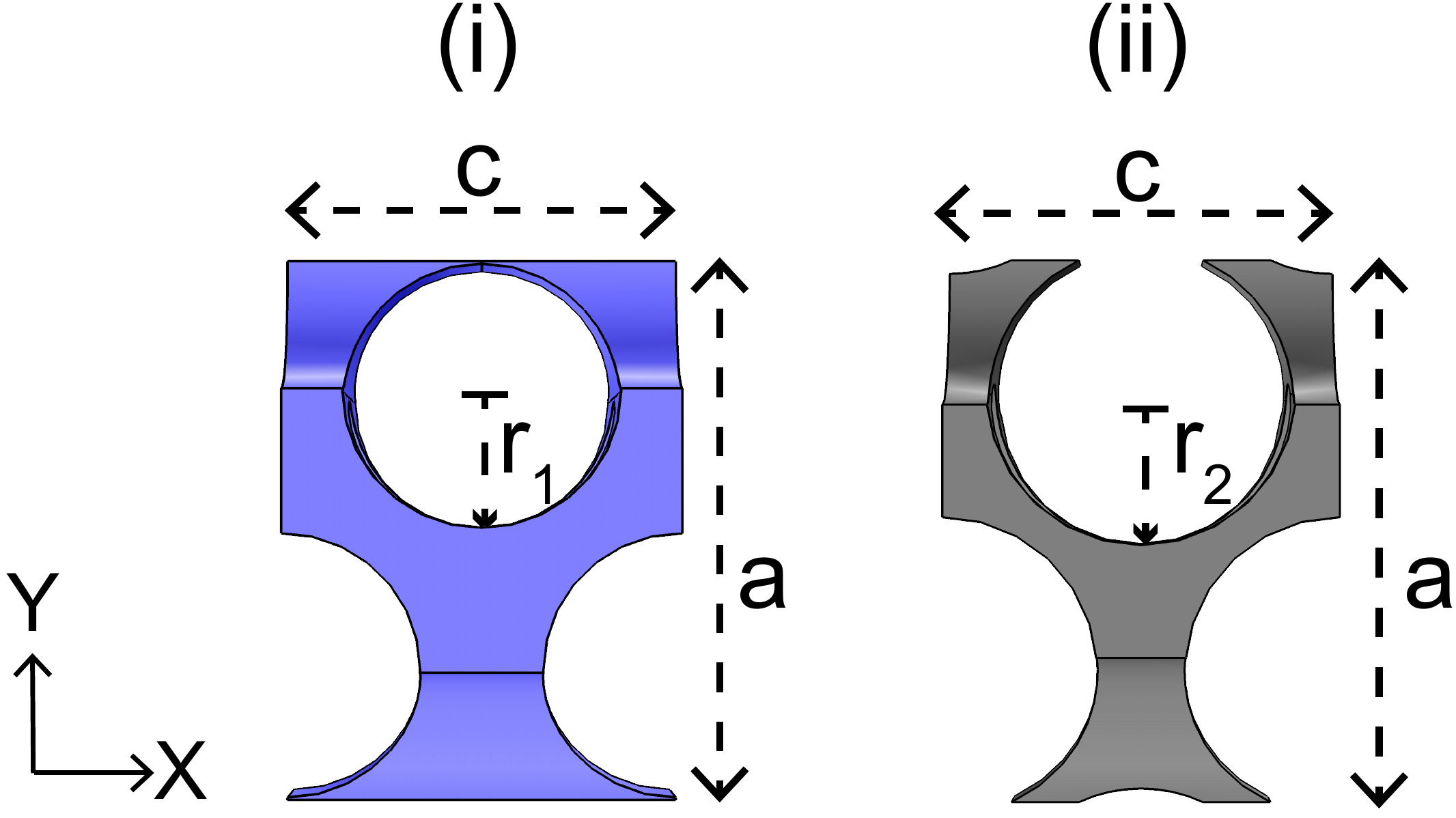}  
        \end{minipage}
        \begin{minipage}[b]{0.70\textwidth}   
            \centering 
            \includegraphics[width=\textwidth]{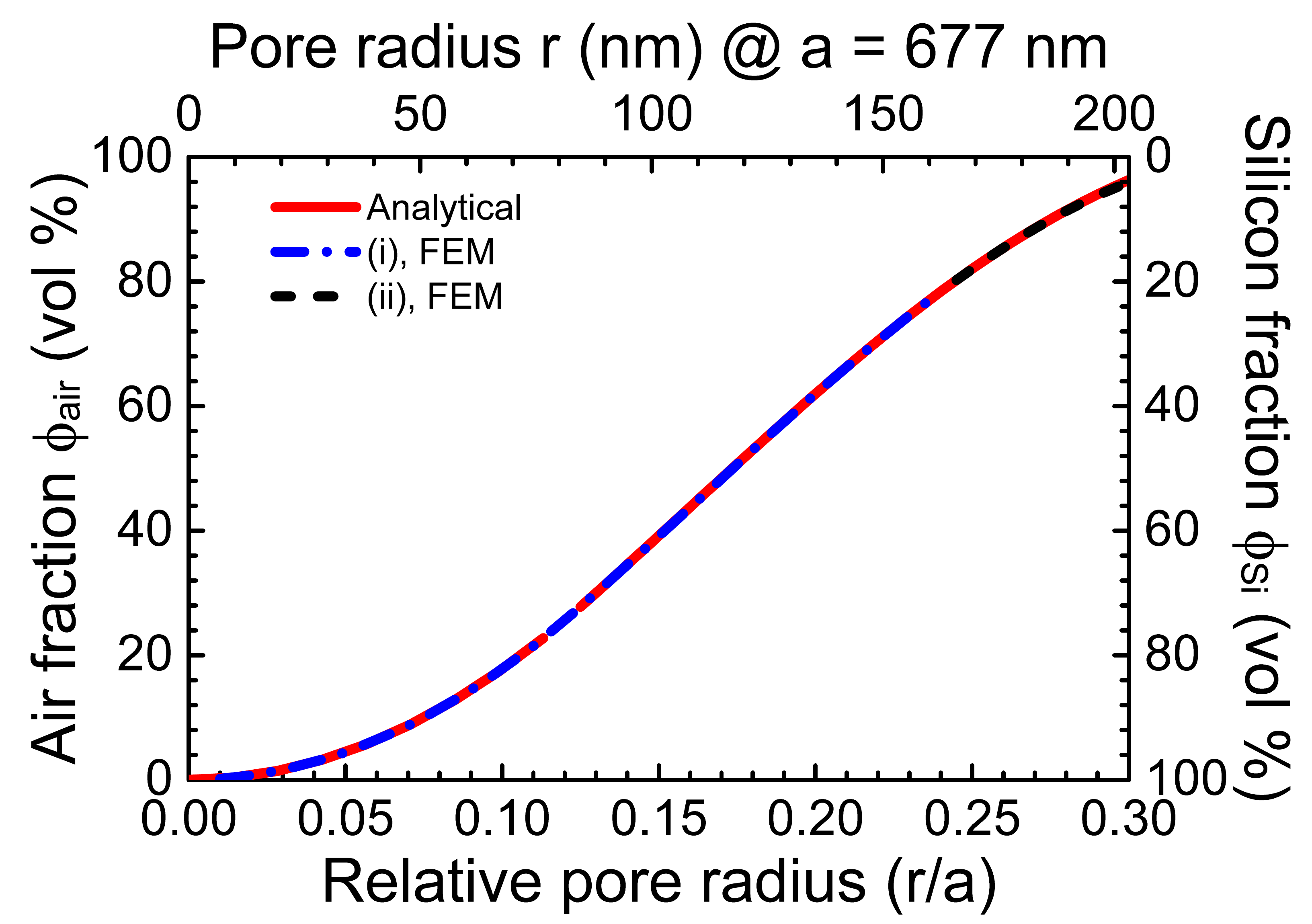} 
        \end{minipage}
        \caption{Top: (i) The tetragonal primitive unit cell of the cubic inverse woodpile photonic crystal structure along the $\mathrm{Z}$ axis with lattice parameters $c$ and $a$ and the pore radius $\frac{r_{1}}{a} = 0.245$, (ii) unit cell adapted to a larger pore radius $\frac{r_{2}}{a} = 0.275$. The blue and black colors in (i) and (ii), respectively, indicate the high-index backbone of the crystal. The white color represents air. 
        Bottom: Volume fraction of air in the 3D inverse woodpile photonic crystal versus the relative pore radius $\frac{r}{a}$. 
        The blue dashed-dotted curve indicates the numerical result for a pore radius between $\frac{r}{a} = 0$ and $\frac{r}{a} = 0.245$ using the primitive unit cell in (i). 
        The black dashed curve indicates the numerical result for a pore radius between $\frac{r}{a} = 0.245$ and $\frac{r}{a} = 0.30$ using the modified unit cell in (ii). 
        The red solid curve represents previously unpublished analytical results by Femius Koenderink (2001).}
        \label{fig:VolumeFraction}
    \end{figure}

For a cubic inverse woodpile with lattice parameters $c$ and $a$, Fig.~\ref{fig:VolumeFraction} (i) shows the tetragonal primitive unit cell for reduced nanopore radii $\frac{r_{1}}{a} = 0.245$. 
This unit cell is periodic in all three directions $\mathrm{X, Y, Z}$. 
If the air volume fraction is further increased by increasing the nanopore radii, Fig.~\ref{fig:VolumeFraction} (ii) reveals remarkable crescent-like shapes appearing at the front and the back interfaces in the $\mathrm{XY}$ view of the unit cell, here for reduced pore radii $\frac{r_{1}}{a} = 0.275$. 
Once the pore radii exceed $\frac{r}{a} \geq 0.245$, the adjacent pores intersect with each other and hence these crescent-like shapes occur as they preserve the periodicity of the unit cell.

Figure~\ref{fig:VolumeFraction} (bottom) shows the calculated volume fraction of air and silicon in the inverse woodpile crystal structure versus the reduced nanopore radius $\frac{r}{a}$ by employing a volume integration routine of the finite element method~\cite{COMSOLMultiphysics}. 
To preserve periodicity of the numerically approximated unit cell, we consider the primitive unit cell in (i) for a pore radius between $\frac{r}{a} = 0$ and $\frac{r}{a} = 0.245$ and the modified unit cell in (ii) for a pore radius between $\frac{r}{a} = 0.245$ and $\frac{r}{a} = 0.30$. 
Our numerical calculation agrees to great precision (within about $10^{-6}\%$) with the analytical results for all pore radii. 
Since an inverse woodpile crystal consists of nearly $80 \%$ air by volume fraction at the optimal pore radius $\frac{r}{a} = 0.245$, it is a very lightweight component for light-absorbing high-tech devices (including photovoltaics), in comparison to bulk silicon with the same thickness.  

\section{Brillouin zone of the 3D inverse woodpile photonic crystal structure}
\label{sect:Brillouinzone}

\begin{figure}[ht]
\centering
\includegraphics[width=0.35\columnwidth]{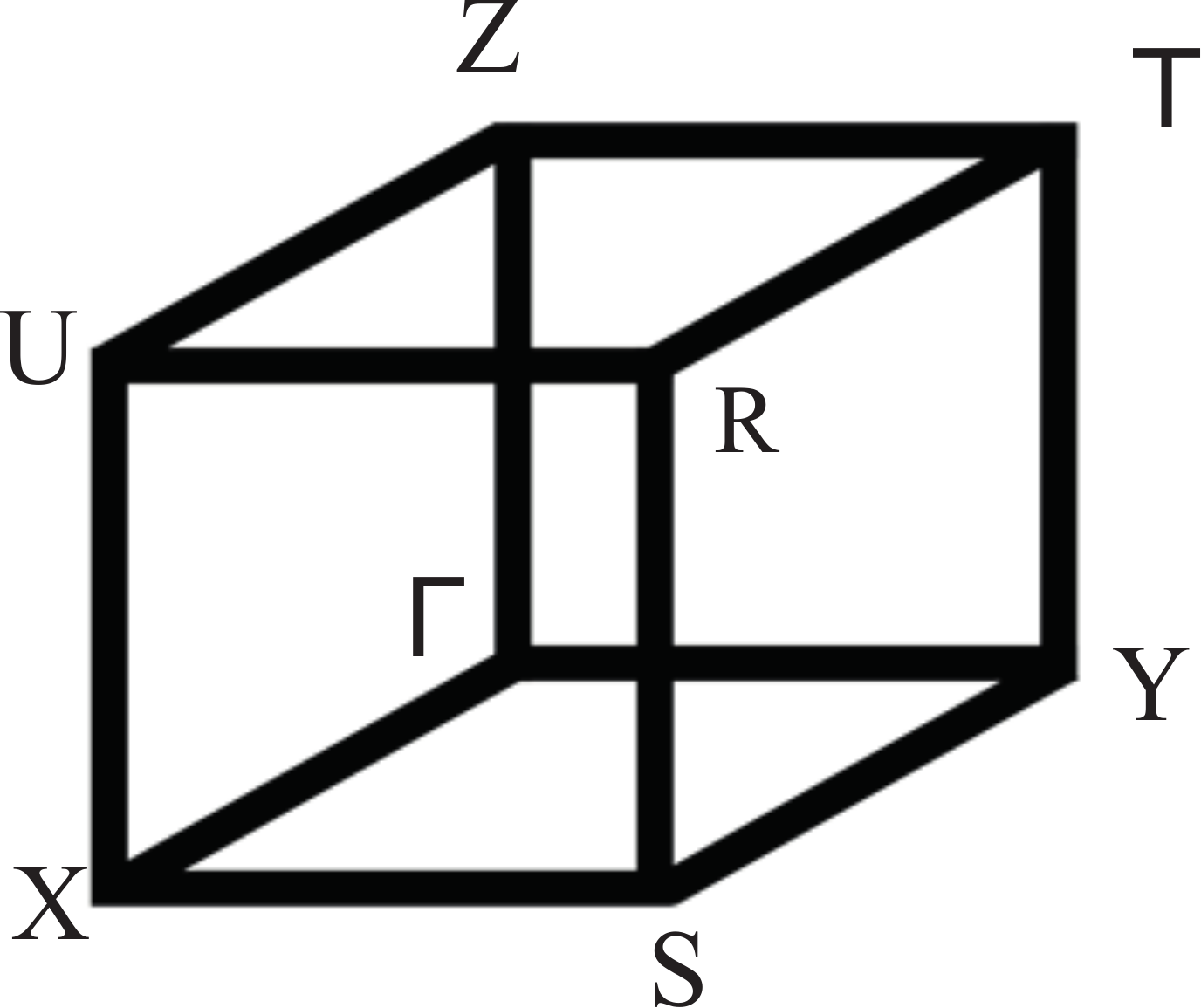}
\caption{
First Brillouin zone of the inverse-woodpile crystal structure in the tetragonal representation showing the high-symmetry points (Roman symbols) and the origin at $\Gamma$. } 
\label{fig:Brillouinzone}
\end{figure}
Figure~\ref{fig:Brillouinzone} shows the first Brillouin zone of the inverse-woodpile crystal structure in the representation with a tetragonal unit cell, with real space lattice parameters $\mathrm{(a, c, a)}$, and reciprocal space lattice parameters  $\mathrm{(2\pi / a, 2\pi / c, 2\pi / a)}$. 
Eight high symmetry points are shown, where $\mathrm{\Gamma}$ corresponds to coordinates (0,0,0), $\mathrm{X}$ to $(1/2, 0, 0)$, $\mathrm{Y}$ to $(0, 1/2, 0)$, and $\mathrm{Z}$ to $(0, 0, 1/2)$. 
The $\mathrm{X}$ and $\mathrm{Z}$ directions correspond to the directions parallel to the two sets of nanopores in the crystal structure, that turn out to be symmetry equivalent, see Refs.~\cite{Huisman2011PRB, Devashish2017PRB}.

\section{Complex and dispersive refractive index of silicon}
\label{sect:RefractiveIndex_Si}
\begin{figure}[ht]
\centering
\includegraphics[width=0.75\columnwidth]{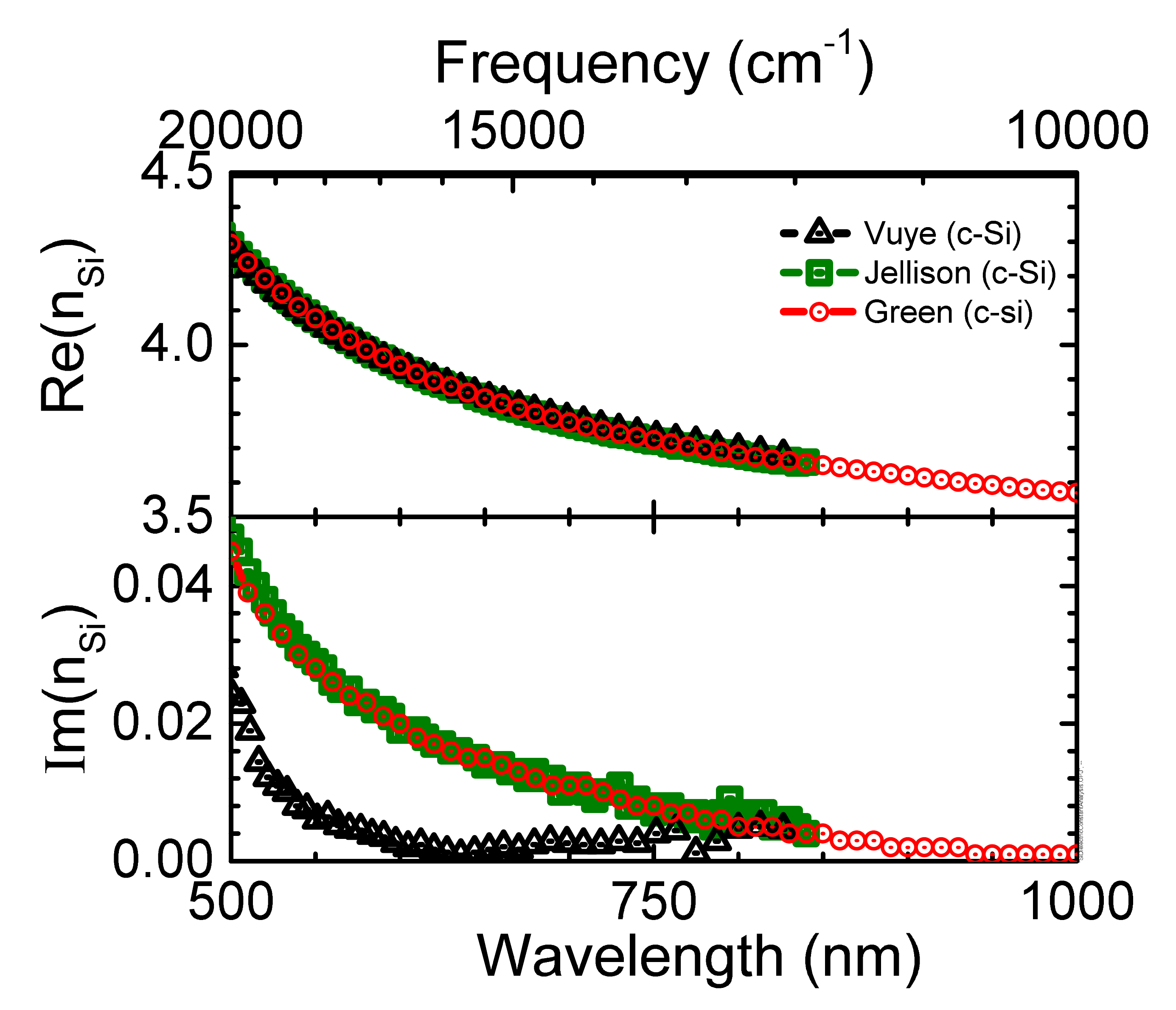}
\caption{Wavelength dependence of the real and imaginary parts of the refractive index of crystalline silicon (c-Si) in the visible and near infrared spectral ranges. 
Red circles, black triangles, and green squares are the data obtained from Ref.~\cite{Green2008SolarEnergyMater}, Ref.~\cite{Vuye1993ThinSolidFilms}, and Ref.~\cite{Jellison1992OptMater}, respectively. 
The top ordinate shows the frequency in wave numbers (cm$^{-1}$). } 
\label{fig:nSi}
\end{figure}

To make our study relevant to practical devices, we employ the dispersive and complex refractive index obtained from experiments to model silicon in all thin films and in the photonic crystal backbone. 
Figure~\ref{fig:nSi} shows the wavelength dependency of the real and imaginary parts of the refractive index of silicon in the visible regime from several sources~\cite{Vuye1993ThinSolidFilms, Jellison1992OptMater, Green2008SolarEnergyMater}. 
Vuye \textit{et al.} report the dielectric function of a commercially available silicon wafer using \textit{in situ} spectroscopic ellipsometry~\cite{Vuye1993ThinSolidFilms}. 
Jellison measured the dielectric function of crystalline silicon using two-channel polarization modulation ellipsometry~\cite{Jellison1992OptMater}, and Green gives a tabulation of the optical properties of intrinsic silicon based on many different sources, aiming at solar cell calculations~\cite{Green2008SolarEnergyMater}. 
Figure~\ref{fig:nSi} shows that for the real part of the refractive indices of Refs.~\cite{Vuye1993ThinSolidFilms},~\cite{Jellison1992OptMater}, and~\cite{Green2008SolarEnergyMater} are in very good mutual agreement, and are thus used in our simulations. 
For the imaginary part of the refractive index, we note that the results of Ref.~\cite{Jellison1992OptMater} and~\cite{Green2008SolarEnergyMater} agree well with each other between $\lambda = 500$ nm and 750 nm and differ from the ones from Ref.~\cite{Vuye1993ThinSolidFilms} for reasons unknown to us. 
All data sets agree well beyond $\lambda = 750$ nm. 
Since Ref.~\cite{Green2008SolarEnergyMater} is based on many different sources of data, we have chosen to adopt it as the imaginary refractive index of silicon in our study.

To assess the impact of adding doped layer, a doping level of $10^{16}$ $\textrm{cm}^{-3}$ already gives substantial mobility. 
Fortunately, from the well-known semiconductor data collection of the Ioffe Institute ("SVM") we find that this doping level yields a very small change in optical properties ($\mathbb{R}\mathrm{e}(n_{Si})$, $\mathbb{I}\mathrm{m}(n_{Si})$). 
Hence, the variation of intrinsic absorption versus doping level is below 1\%~\cite{Wolfson1967FizTekhPol}, and thus for practical purposes negligible.

\section{Photonic band gap generated current density}
\label{sect:JSC}

\begin{table}[]
    \centering
    \begin{tabular}{|c|c|c|}
        \hline
       $J_{PBG}$  & $s-$stop band & $p-$stop band \\
       $(mA/cm^{2})$ & ($\lambda_{\mathrm{up}} = 640$ nm to $\lambda_{\mathrm{low}} = 890$ nm) & ($\lambda_{\mathrm{up}} = 650$ nm to $\lambda_{\mathrm{low}} = 870$ nm)\\
       \hline
       \hline
       $L_{Si} = 80$ nm Si only &  0.2244 & 0.1901 \\
       \hline
       $L_{Si} = 80$ nm Si + perfect metal &  0.1728 & 0.1544  \\
       \hline
       $L_{Si} = 80$ nm Si + 3D photonic crystal &  2.9442 & 2.0362\\
       \hline
       $L_{Si} = 80$ nm Si, Lambertian scattering & 6.1607 & 5.5523 \\
       \hline
    \end{tabular}
    \caption{Photonic band gap generated current density $J_{PBG}$ computed using Eq.~\ref{eq:ShortCircuitCurrentDensity} for different ultrathin silicon film devices: thin film only, with a perfect metal back reflector, with a 3D photonic band gap back reflector, and with Lambertian scattering. 
    }
    \label{tab:JSC_Sub}
\end{table}

In an ideal situation, where one absorbed photon generates one electron-hole pair, it is known from the solar cell literature) that the wavelength-dependent external quantum efficiency $EQE(\lambda)$ is equal to the absorption $A_{Si}(\lambda)$~\cite{Luque2011Book, Massiot2020NatEnergy}. 
Hence, we define a current density $J_{PBG}$ generated by incident light within the photonic band gap (PBG)  
\begin{equation}
J_{PBG} \equiv q \int^{\lambda_{\mathrm{low}}}_{\lambda_{\mathrm{up}}} EQE(\lambda)~{P_{AM 1.5}(\lambda)}~d\lambda, 
\label{eq:ShortCircuitCurrentDensity}
\end{equation}
where $\lambda_{\mathrm{low}}$ is the wavelength corresponding to the lower band edge of the band gap, $\lambda_{\mathrm{up}}$ the wavelength corresponding to the upper band edge of the band gap. 
In other words, Eq.~\ref{eq:ShortCircuitCurrentDensity} is an integral of the absorption spectrum weighted with the solar spectrum of incident photons. 

\begin{table}[]
    \centering
    \begin{tabular}{|c|c|c|}
        \hline
       $J_{PBG}$  & $s-$stop band & $p-$stop band \\
       $(mA/cm^{2})$ & ($\lambda_{\mathrm{up}} = 640$ nm to $\lambda_{\mathrm{low}} = 890$ nm) &  ($\lambda_{\mathrm{up}} = 650$ nm to $\lambda_{\mathrm{low}} = 870$ nm) \\
       \hline
       \hline
       $L_{Si} = 2400$ nm Si only & 4.1635 & 3.6452 \\
       \hline
       $L_{Si} = 3600$ nm Si only & 5.1025 & 4.7038 \\
       \hline
       $L_{Si} = 2400$ nm + 1200 nm 3D photonic crystal & 9.2394 & 8.9276\\
       \hline
       $L_{Si} = 2400$ nm, Lambertian scattering & 16.2197 & 14.4055 \\
       \hline
       $L_{Si} = 3600$ nm Si, Lambertian scattering & 16.6290 & 14.7370 \\
       \hline
    \end{tabular}
    \caption{Photonic band gap generated current density $J_{PBG}$ calculated using Eq.~\ref{eq:ShortCircuitCurrentDensity} at normal incidence for thin silicon films only ($L_{Si} = 2400$ nm and $L_{Si} = 3600$ nm), with a 3D inverse woodpile photonic crystal back reflector, and with Lambertian scattering. 
    }
    \label{tab:JSC_Supra}
\end{table}

Tables~\ref{tab:JSC_Sub} and~\ref{tab:JSC_Supra} show the photonic band gap generated current density $J_{PBG}$ for normal incidence  (Eq.~\ref{eq:ShortCircuitCurrentDensity}) for both the supra-wavelength $L_{Si} = 2400$ nm thin and sub-wavelength $L_{Si} = 80$ nm ultrathin silicon films, and with and without various back reflectors. 
Table~\ref{tab:JSC_Sub} shows that the perfect metal back reflector does not increase the photonic band gap generated current density $J_{PBG}$ for an ultrathin film thickness much less than the wavelength in the material ($L_{Si} << \lambda/n$). 
This result confirms our observation in Sec.~\ref{subsec:Subwavelength} that a perfect metal hardly enhances absorption for films with thicknesses corresponding to a Fabry-P\'erot minimum for the desired wavelengths of enhancement. 

In presence of a 3D photonic band gap back reflector, the photonic band gap generated current density $J_{PBG}$ for the $L_{Si} = 80$ nm ultrathin film increases nearly $13\times$ for the $s-$polarized stop band and $10\times$ for the $p-$polarized stop band, which supports the absorption enhancements observed for 3D photonic crystal back reflector in Sec.~\ref{subsec:ComparisonPerfectMetal}. 
Moreover, similar to absorption enhancement results in Sec.~\ref{subsec:Subwavelength} the 3D photonic band gap back reflector also increases the photonic band gap generated current density $J_{PBG}$ for a $L_{Si} = 2400$ nm thin film by nearly $2.21\times$ for $s-$stop band and $2.45\times$ for $p-$stop band (Table~\ref{tab:JSC_Supra}). 
Finally, our computation in Table~\ref{tab:JSC_Supra} supports the use of a photonic crystal back reflector by increasing the photonic band gap generated current density $J_{PBG}$ by $1.81\times$ for $s-$stop band and $1.9\times$ for $p-$stop band even compared to a $L_{Si} = 3600$ nm thin film (similar to absorption discussion in Sec.~\ref{subsect:backbone}). 

On the other hand, both Table~\ref{tab:JSC_Sub} and Table~\ref{tab:JSC_Supra} show that the photonic band gap generated current density $J_{PBG}$ with a photonic crystal back reflector is still nearly $2-2.5\times$ lower than the one computed for Lambertian scattering. 
This difference is attributed to the fact that absorption due to Lambertian scattering also assumes a perfect anti-reflection front coating, unlike our present computations, see~ Fig~\ref{fig:ComputationalCell}.



\begin{thebibliography}{99}
\bibitem{ComputationalMesh}
Since the air layer and the thin silicon film are homogeneous media, an upper limit $\triangle l \leq \frac{\lambda_{0}}{8}$ is imposed on the edge length $\Delta l$ of any tetrahedron in these layers, with $\lambda_{0}$ the shortest wavelength of the incident plane waves in vacuum. 
Since an inverse woodpile crystal contains many sharp and curved interfaces between the high-index backbone and the low-index medium, an upper limit $\Delta l \leq \frac{\lambda_{0}}{8.\max(n_{Si})}$ is imposed on the edge length $\Delta l$ of any tetrahedron inside the photonic crystal (as well as a sub-wavelength thin silicon film $L_{Si} = 80$ nm), where $\max(n_{Si})$ is the maximum silicon refractive index in the relevant spectral range.
\bibitem{Serendipity}
"Serendipity" is a high performance computing cluster whose main features are 16 Dell Power edge R430 servers with each 2x Intel Xeon E5-2698 processors at 2.2GHz, with in total 640 cores and about 3.6 terabyte memory, connected through Infiniband. 
The head node is a Dell Power Edge R730.
\bibitem{DiffWithLRSPP}
There is an important difference with a LRSPP, namely that here the field is maximal at the center of the thin silicon film, whereas LRSPP fields are minimal at the center of a thin metal film.
\end{thebibliography}
\end{document}